\documentclass[10pt]{IEEEtran}
\IEEEoverridecommandlockouts
\normalsize

\ifCLASSINFOpdf

\else
\usepackage{graphicx}
\usepackage{subcaption}
\fi
\usepackage{soul}
\usepackage{array}
\usepackage{color}
\usepackage{amsfonts}
\usepackage{amssymb}
\usepackage{amsmath}
\usepackage[pdftex]{graphicx}
\usepackage{cite}
\usepackage{amsthm}
\theoremstyle{remark} 

\usepackage{balance}
\usepackage{caption}
\usepackage{subcaption}
\usepackage{textcomp}
\usepackage{gensymb}
\usepackage{float}
\usepackage{comment}
\usepackage{subcaption}
\usepackage{tabularx}
\usepackage{multirow}
\usepackage{amsthm}
\usepackage{url}
\usepackage{booktabs}
\usepackage{algorithm}
\usepackage{tikz}
\usetikzlibrary{arrows.meta,positioning,calc,fit,shapes.geometric,shapes.misc}

\usepackage[noend]{algpseudocode}
\makeatletter
\renewcommand{\ALG@beginalgorithmic}{\footnotesize}
\makeatother

\usepackage[utf8]{inputenc}

\IEEEoverridecommandlockouts
\usepackage{fancyhdr}
\fancypagestyle{IEEEtitlepagestyle}{
    \fancyhf{} 
    \fancyhead[C]{Submitted for possible publication to IEEE. Paper currently under review. The contents of this paper may change at any time
without notice.}
    
}

\begin{document}
\title{Digital-Twin Empowered Deep Reinforcement Learning For Site-Specific Radio Resource Management in NextG Wireless Aerial Corridor}

\author{Pulok Tarafder,~\IEEEmembership{Graduate Student Member,~IEEE}, Zoheb Hassan,~\IEEEmembership{Member,~IEEE}, Imtiaz Ahmed,~\IEEEmembership{Senior Member,~IEEE}, Danda B. Rawat,~\IEEEmembership{Senior Member,~IEEE}, Kamrul Hasan,~\IEEEmembership{Senior Member,~IEEE}, Cong Pu, ~\IEEEmembership{Member,~IEEE,}

\thanks{P. Tarafder, I. Ahmed, and D. B. Rawat are with Dept. of Electrical Engineering and Computer Science, Howard University, Washington, DC 20059, USA (email: pulok.tarafder@bison.howard.edu;  imtiaz.ahmed@howard.edu; danda.rawat@Howard.edu).}
\thanks{Z. Hassan is with Dept. of Electrical and Computer Engineering, Université Laval, Québec, G1V 0A6, Canada (email: md-zoheb.hassan@gel.ulaval.ca).}
\thanks{K. Hasan is with Dept. of Electrical and Computer Engineering, Tennessee State University, Nashville, TN 37209, USA  (email: mhasan1@tnstate.edu).}
\thanks{C. Pu is with Dept. of Computer Science, Oklahoma State University, Stillwater, OK 74078 USA (e-mail: cong.pu@outlook.com).}

\thanks{This work was supported in part by the NSF Grant \# 2200640, in part by DoD/US Army Contract W911NF-22-1-022, and in part by the US DoD Center of Excellence in AI/ML at Howard University under Contract W911NF-20-2-0277 with the US ARL. However, the views and conclusions expressed herein are those of the authors and do not necessarily represent the official policies of the funding agencies.}}



\maketitle

\begin{abstract}
Joint base station (BS) association and beam selection in multi unmanned aerial vehicle (UAV) aerial corridors constitutes a challenging radio resource management (RRM) problem. It is driven by high-dimensional action spaces, need for substantial overhead to acquire global channel state information (CSI), rapidly varying propagation channels, and stringent latency requirements. Conventional combinatorial optimization methods, while near-optimal, are computationally prohibitive for real-time operation in such dynamic environments. While learning-based approaches can mitigate computational complexity and CSI overhead, the need for extensive site-specific (SS) datasets for model training remains a key challenge. To address these challenges, we develop a Digital Twin (DT)-enabled two-stage optimization framework that couples physics-based beam gain modeling with deep reinforcement learning (DRL) for scalable online decision-making. In the first stage, a channel twin (CT) is constructed using a high-fidelity ray-tracing solver with geo-spatial contexts, and network information to capture SS propagation characteristics, and dual annealing algorithm is employed to precompute optimal transmission beam directions. In the second stage, a Multi-Head Proximal Policy Optimization (MH-PPO) agent, equipped with a scalable multi-head actor–critic architecture, is trained on the DT-generated channel dataset to directly map complex channel and beam states to jointly execute UAV-BS-beam association decisions. The proposed PPO agent achieves a 44\%-121\% improvement over Deep Q Network (DQN) and 249\%-807\% gain over traditional heuristic based optimization schemes in a dense UAV scenario, while reducing inference latency by several orders of magnitude. These results demonstrate that DT-driven training pipelines can deliver high-performance, low-latency RRM policies tailored to SS deployments, thereby establishing a viable pathway for real-time resource management in next-generation aerial corridor networks.
\end{abstract}

\begin{IEEEkeywords}
Digital-Twin, UAVs, Resource Allocation, Deep Reinforcement Learning, Multi-Head PPO.
\end{IEEEkeywords}

\section{Introduction}

Unmanned aerial vehicles (UAVs) have emerged as a key research pivot in sixth-generation (6G) wireless communication systems. Future UAV-related studies are expected to be driven by advances in intelligent software and hardware platforms, including graphical processing units (GPUs), digital twins (DTs), and machine learning (ML) techniques \cite{Lin2023DT}. Multiple UAVs, hereafter referred to collectively as drones with no human pilot on board, have attracted sustained attention due to their agility, autonomy, and suitability for diverse missions such as search and rescue, target tracking, forest fire prevention, crowd monitoring, and agricultural spraying \cite{beard2012small, kuang2019deep, atif2021uav, xiao2021blockchain, singh2020trajectory}. However, due to limited energy reserves and payload capacity, a single UAV is often insufficient for complex forthcoming 6G tasks. For a multi-UAV system with a dynamic and uncertain complex environment, understanding the control behavior and making system-wide decision is strenuous \cite{cattai2025multi}. These limitations have motivated the integration of multi-UAV cooperation with DT technology to enhance network-wide task performance \cite{Tang2023DTUAV}.

Next-generation UAV-enabled wireless systems are envisioned to be highly autonomous, with capabilities for self-optimization, self-debugging, and self-configuration \cite{khan2022digital}. Advanced artificial intelligence (AI) techniques can dynamically adapt to changing network conditions and user demands. However, their long training times and large data requirements make them unsuitable for mission-critical, real-time 6G applications. To provide ultra-reliable, high-rate connectivity across diverse services, 6G networks must adopt proactive, online learning that can rapidly adapt to heterogeneous architectures and evolving user behavior. Conventional simulators, long used across the physical, medium access control (MAC), and network layers, are limited to offline, scenario-specific analyses and lack real-time feedback. Moreover, the conventional link-level Monte Carlo simulators rely on statistical channel models for performance analyses that may not accurately capture the real-world scenario for diversified use cases.

The concept of Digital Twins (DTs), initially introduced by Grieves and later formalized by Grieves and Vickers in \cite{grieves2016digital}, extends far beyond traditional simulation. Unlike simulators that capture isolated processes, DTs maintain a continuous, two-way exchange of information with physical systems, enabling real-time monitoring, predictive analytics, and closed-loop control. DTs were originally applied to manufacturing, however, recently they have been expanded into domains such as healthcare, transportation, smart cities, and robotics \cite{mihai2022digital}, and more recently have been recognized as a transformative enabler for wireless networks and 6G systems \cite{tao2024wireless}.

In the context of UAV, DT environments provide physics-consistent channel realizations, mobility-aware models, and accurate beam-management feedback for evaluating throughput-maximization and link-adaptation strategies prior to live flight tests. DT platforms further enable end-to-end emulation of system behavior, yields near-real-world performance metrics and supports iterative refinement of next generation (NextG) algorithms within a controllable virtual space. These virtual models can remain synchronized with physical networks, delivering site-specific (SS) channel realizations and network parameters while also enabling exploratory ``what-if" analyses \cite{Haider2025DTPrecoding,Elloumi2025DTInterference}. Advanced capabilities such as city-scale three-dimensional (3D) mapping, multi-modal sensing integration, and low-latency inference for real-time reconfiguration further position DTs as a key enabler for NextG communications \cite{Alkhateeb2023RTDT}. Moreover, large-scale, physically grounded DT replicas support efficient radio-access-network (RAN) planning by allowing virtual evaluation of base-station (BS) placements, antenna configurations, and resource-management policies. This reduces the cost, time, and risk associated with field-trial-driven deployments.
\vspace{-1.0em}
\subsection{Related Works, Motivation, and Contribution}
Recent studies in the literature underscore the growing synergy between deep reinforcement learning (DRL) and DT frameworks for wireless communications. For instance, Tang $\emph{et al.}$ \cite{Tang2023DTUAV} developed a DT-driven DRL framework for task assignment in multi-UAV systems, while Karimi-Bidhendi $\emph{et al.}$ \cite{Karimi2024UAVCorridor} studied the optimization of cellular networks for UAV corridors using quantization theory. Further work includes Haider $\emph{et al.}$'s \cite{Haider2025DTPrecoding} DT-enabled channel precoding method based on over-the-air channel impulse response (CIR) inference and Elloumi $\emph{et al.}$'s \cite{Elloumi2025DTInterference} proactive DT-driven interference management strategies for vehicular networks. Collectively, while these studies highlight the potential of DT and DRL methods, they fall short of addressing the joint UAV-BS association and beam selection problem in corridor-style UAV networks. Beyond these specific applications, Lin $\emph{et al.}$ \cite{Lin2023DT} and Alkhateeb $\emph{et al.}$ \cite{Alkhateeb2023RTDT} highlighted DT as foundational enablers for 6G, emphasizing their role in real-time optimization and predictive control. Specifically in the UAV domain, the capabilities demonstrated by Tang $\emph{et al.}$ \cite{Tang2023DTUAV} in task allocation and Karimi-Bidhendi $\emph{et al.}$'s \cite{Karimi2024UAVCorridor} introduced the concept of structured UAV corridors and explored cellular optimization in aerial highways. Advances in SS channel modeling and interference management using DT-driven methods \cite{Haider2025DTPrecoding, Elloumi2025DTInterference} further confirm this trend. Therefore, these studies collectively motivate the adoption of DRL within DT-powered UAV corridor networks, where high-dimensional channel states, dynamic UAV mobility, and stringent latency requirements demand scalable and adaptive learning frameworks. 

Although interest in DT-enabled aerial communication has grown substantially, a unified framework that jointly optimizes UAV-BS association and beam selection in a scalable and adaptive manner remains nonexistent. Addressing this gap requires the development of an accurate digital replica of the wireless channel, a component known as the \emph{Channel Twin (CT)}. A precise CT is foundational to robust resource allocation, as it provides the physics-consistent channel state information which is required for reliable NextG algorithm design and performance validation.

Traditional optimization techniques are generally effective for static resource-allocation problems. However, their reliance on global channel state information (CSI) and iterative search renders them computationally prohibitive in the highly dynamic environments of 6G aerial corridor, where channel conditions evolve rapidly and as a result, significant overhead for frequent CSI collection and real-time adaptation are essential. To overcome these computational and adaptivity challenges, we advocate a paradigm shift toward intelligent, learning-based resource management. DRL provides a powerful methodology for network entities to acquire near-optimal control policies directly through environmental interaction. However, DRL convergence relies on extensive exploration, which can cause significant disruptions to network operations. Therefore, trustworthy offline DRL training is essential. This work addresses this challenge by proposing a novel DRL framework that leverages a high-fidelity CT to train an agent for joint UAV–BS association and beam selection, with the objective of mitigating interference and maximizing the network sum rate in NextG aerial corridor systems. The specific contributions are summarized as follows:

1) \textbf{DT-driven training methodology:} We construct a physics-consistent offline dataset using a high-fidelity CT that integrates geo-spatial features, BS, UAV locations, and SS ray tracing, conforming to Electronic Communications Committee (ECC) Report~281 \cite{ECCReport281} and 3rd Generation Partnership Project (3GPP) TR~37.840 \cite{3gpp-tr-37840} specifications. This dataset underpins the training of a Proximal Policy Optimization (PPO) agent; an on-policy, policy-gradient DRL algorithm designed for environments with discrete or continuous action spaces, thereby enabling robust policy learning across diverse channel conditions and network topologies.

2) \textbf{PPO-based resource-allocation framework:} A coordinated UAV–BS–beam association problem is formulated to maximize the network sum capacity in the presence of inevitable inter-cell interference. Since this optimization problem is computationally intractable and NP-hard, it is reformulated as a policy-learning task. A multi-head PPO (MH-PPO) agent is trained to directly map high-dimensional CSI features to optimal resource-allocation decisions, eliminating the need for computationally expensive iterative optimization during online deployment.

3) \textbf{Ensuring Scalability and Load-balancing:} A specialized multi-head actor-critic neural network architecture is proposed in which a shared trunk processes global state information, while dedicated actor heads produce actions for individual UAVs. This design inherently scales with varying UAV densities.  Meanwhile, MH-PPO framework enforces per-BS capacity constraints through a novel penalty-based reward mechanism rather than hard optimization constraints. When multiple UAVs select the same BS, the environment admits the top-$N$ strongest links and penalizes excess UAVs, enabling the agent to learn load-balanced associations without explicit constraint encoding in the policy network. This approach achieves fairness through learned behavior rather than algorithmic enforcement, maintaining real-time inference capability.

5) \textbf{Superior performance with real-time inference:} Extensive simulations, conducted over a realistic 3D virtual environment with 3GPP-standarized antenna parameters, demonstrate that the proposed DT-enabled MH-PPO framework achieves significantly higher average throughput compared with optimization-based baselines. It delivers inference latencies several orders of magnitude lower, validating its suitability for real-time 6G UAV corridor operations.  
\vspace{-10pt}
\subsection{Relation to Prior Work and Paper's Organization}
This paper extends our earlier conference contribution \cite{tarafder2025digital}, which presented a DT-guided Hungarian optimization framework for activating UAV-BS-beam association that maximizes signal strength, without considering inter-cell interference. The present work addresses this limitation by proposing a DT-enabled PPO optimization framework that explicitely accounts for inter-cell interference in the reward design, thereby improving solution optimality while enhancing scalability for near-real-time decision making. Furthermore, we provide new baseline comparisons, a detailed environment design, and a comprehensive performance evaluation, thereby establishing the first DT-powered DRL benchmark for UAV corridor communications. 

The remainder of this paper is organized as follows. In Section \ref{sec:system_model} we present the system model. The problem formulation and our proposed MH-PPO based resource allocation framework is formulated in Section \ref{sec:ppo_framework}. In Section \ref{sec:results}, we present simulation results of our proposed DT framework. Finally, we conclude our work in Section \ref{sec:conclusion}.

\begin{figure}[h]
\centering
\includegraphics[scale=.175]{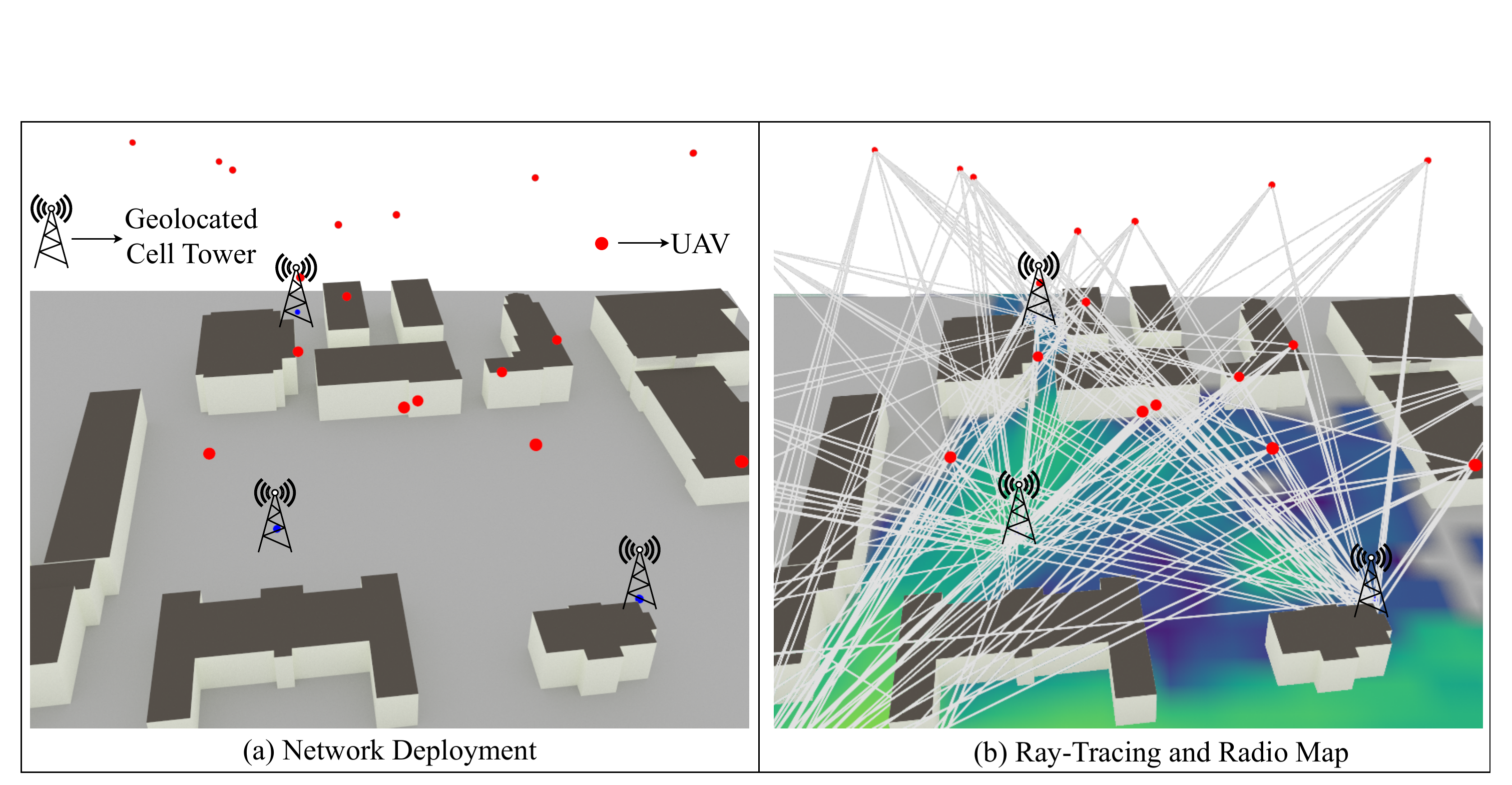}
\caption{UAVs and BSs in the site-specific (SS) DT environment.}
\label{fig:Howard DT Scene}
\end{figure}
\vspace{-1.0em}
\section{System Model}
\label{sec:system_model}
We consider a DT-enabled aerial corridor network comprising a set of $M$ UAVs, $\mathcal{M}=\{1, \dots, M\}$, acting as receivers, and a set of $L$ terrestrial BSs, $\mathcal{L}=\{1, \dots, L\}$. Each BS is equipped with a uniform planar array (UPA) antenna capable of generating a set of $N$ orthogonal beams obtained using 3GPP style codebook, $\mathcal{N}=\{1, \dots, N\}$, where each beam is defined for a particular direction. The total transmit power of each BS is equally distributed among its $N$ beams. Our objective is to learn an association policy in which each UAV is assigned to exactly one BS–beam pair, and each beam serves at most one UAV, and the network sum capacity is maximized while effectively mitigating both inter-cell and intra-cell interference.
\vspace{-1.0em}
\subsection{Channel Twin Construction}
To capture the complex propagation environment, we construct a CT using NVIDIA \textit{Sionna}~\cite{hoydis2023sionna}, a GPU-accelerated, open-source link-level simulator that integrates a physics-based ray-tracer built on Mitsuba-3 and TensorFlow \cite{tensorflow2015-whitepaper, Mitsuba3}. The workflow consists of three stages: 

1) \textit{3D Environment Modeling}: Without loss of a generality, a SS model of the Howard University, Washington, DC, USA campus with coordinates [38.92490, -77.02254] for the northwest and [38.92184, -77.01850] for the southeast is created using OpenStreetMap building footprints imported into Blender. The scene encompasses 14 buildings with realistic 3D geometries, roads, and parking lot structures. We also incorporated realistic material properties using International Telecommunication Unit (ITU) material profiles, i.e., $\text{itu\_concrete, itu\_marble, itu\_metal}$ to corresponding surfaces so that we obtain realistic reflection diffraction, and penetration characteristics.

2) \textit{Infrastructure Integration}: In this stage, we geolocated cellular BS coordinates from the OpenCelliD~\cite{opencellid} database and embedded into the scene. Without loss of generality, we consider four BSs with the following coordinates: BS$_1$ = [-127, 92, 5], BS$_2$ = [-30, 30, 5.2], BS$_3$ = [115, 36, 5.5], BS$_4$ = [-60, -83, 5], where [x, y, z] meters (m) in the local scene Cartesian coordinate system. Each BS is equipped with a 4$\times$4 uniform planar array (UPA) configured with $d_H = d_V = 0.5\lambda$ element spacing at carrier frequency $f_c = 3.5$ GHz, where $\lambda$ represents the wavelength. 

Unlike trajectory-based models that confine UAVs to predetermined flight paths, our dataset generation randomizes the UAV positions independently within the 3D corridor for each scenario. 
For each scenario with altitude $\xi \in \{60, 80, 100\}$ m and spatial domain dimension of $[-200, 130] \times [-150, 150]$ m, we sample $M$ UAV coordinates $\{(x_m, y_m, z_m)\}_{m=1}^M$ such that:
\begin{equation}
x_m \sim \mathcal{U}[-200, 130], \quad y_m \sim \mathcal{U}[-150, 150], \quad z_m = \xi.
\end{equation}
This static-positioning approach yields diverse propagation conditions spanning a wide range of link distances, incident angles, and line-of-sight (LOS)/non-LOS states, representative of operational UAV deployments in dense urban environments.
3) \textit{Ray Tracing and Channel Extraction}: In the final stage of CT construction, the annotated scene in which UAVs and BSs are positioned according to the sampled coordinates are loaded into \textit{Sionna} from the previous two stages. Afterwards, \textit{Sionna}'s ray-tracer performs path sampling with $N_{\text{rays}} = 10^6$ rays per BS-UAV link with maximum interaction depth $d_{\text{max}} = 5$, enabling accurate modeling of specular multipath propagation. For each link, \textit{Sionna} outputs a complex-valued (CIR) tensor, $\mathcal{Z} \in \mathbb{C}^{M \times L \times N \times K}$, where $K$ represents the number of delay taps capturing temporal dispersion. The channel gain matrix $\mathbf{H} \in \mathbb{R}_{+}^{M \times L \times N}$ is derived by aggregating multipath power as follows:
\begin{equation}
|h_{m,l,n}|^2 = \sum_{k=1}^{K_{m,l,n}} |\mathcal{Z}_{m,l,n,k}|^2,
\end{equation}
where $\mathcal{Z}_{m,l,n,k}$ denotes the complex coefficient of the $k$-th path between UAV $m \in \mathcal{M}$, BS $l \in \mathcal{L}$, and beam index $n \in \mathcal{N}$. In addition, \textit{Sionna} also provides multipath resolved angle-of-arrival (AoA) in spherical coordinates. The zenith angle $\theta \in [0^\circ, 180^\circ]$ and azimuth angle $\phi \in [0^\circ, 360^\circ]$ are extracted for each path component. Mean AoAs are computed by averaging over all $K$ multipath as follows:
\begin{equation}
\bar{\theta}_{m,l,n} = \frac{1}{K_{m,l,n}} \sum_{k=1}^{K_{m,l,n}} \theta_{m,l,n,k},
\end{equation}
\begin{equation}
\bar{\phi}_{m,l,n} = \frac{1}{K_{m,l,n}} \sum_{k=1}^{K_{m,l,n}} \phi_{m,l,n,k}.
\end{equation}
An illustration of the constructed DT module that produces the CT is presented in Fig.~\ref{fig:Howard DT Scene}. 
\vspace{-1.0em}
\subsection{Antenna Gain Model}
The overall antenna gain is modeled according to 3GPP TR~37.840 specifications~\cite{3gpp-tr-37840}. The total gain in the direction $(\theta, \phi)$, denoted $G_{5 \mathrm{G}}$, is the sum of the element gain $A_E$ and the array gain $A_V$:
\begin{equation}\label{eq:total_gain}
    G_{5 \mathrm{G}}\left(\theta, \phi\right) = A_E\left(\theta, \phi\right) + A_V\left(\theta, \phi\right),
\end{equation}
where all gains are presented in dBi. The directional pattern of a single antenna element, $A_E(\theta,\phi)$, is given by:
\begin{equation}
A_E(\theta,\phi) = G_{E,\text{max}} - \min\{-[A_{E,V}(\theta) + A_{E,H}(\phi)], A_m\},
\end{equation}
with the vertical and horizontal attenuation components defined as:
\begin{align}
A_{E,V}(\theta) &= -\min\left\{12\left(\frac{\theta - 90^\circ}{\theta_{3\text{dB}}}\right)^2, \mathrm{SLA}_V\right\} \\
\text{and}~A_{E,H}(\phi) &= -\min\left\{12\left(\frac{\phi}{\phi_{3\text{dB}}}\right)^2, A_m\right\},~\text{respectively}.
\end{align}
The array gain, $A_V(\theta,\phi)$, which results from digital beamforming, is modeled as:
\begin{equation}
A_V\left(\theta, \phi\right) = 10 \log _{10}\left(\left|\mathbf{V}^H\left(\theta, \phi\right) \mathbf{W}\left(\phi_{\text{scan}}\right)\right|^2\right),
\end{equation}
where $\mathbf{V}\left(\theta, \phi\right)$ is the steering vector and $\mathbf{W}\left(\phi_{\text{scan}}\right)$ is the beamforming vector for a given scan angle $\phi_{\text{scan}}$. For an antenna array with $N_H \times N_V$ elements, the $(u,v)$-th elements of the steering and beamforming vectors are expressed as:
\begin{multline}
[\mathbf{V}(\theta, \phi)]_{u,v} = \exp[\frac{2\pi j}{\lambda} (u-1)d_H \sin\theta \sin\phi\\ + (v-1)d_V \cos\theta)],
\end{multline}
and
\begin{multline}
[\mathbf{W}(\phi_{\text{scan}})]_{u,v} = \frac{1}{\sqrt{N_H N_V}} \exp\Bigg[-j\frac{2\pi}{\lambda}\Big( (u-1)d_H\\ \sin(\phi_{\text{SCAN}}) \cos(\theta_{\text{TILT}}) - (v-1)d_V\sin(\theta_{\text{TILT}}) \Big)\Bigg],
\end{multline}
respectively. Here, $N_H$ and $N_V$ are the number of horizontal and vertical antenna elements, respectively. Moreover, $d_H$ and $d_V$ are the element spacings for horizontal and vertical directions, respectively, $\lambda$ denotes the wavelength, $\phi_{\text{SCAN}}$ represents the horizontal scan angle, and $\theta_{\text{TILT}}$ is the downtilt angle.

\subsection{Interference and Throughput Model}

We define the system state by two sets of binary indicator variables. Let $\beta_{m,l} \in \{0,1\}$ be 1 if UAV $m$ is associated with BS $l$, and 0 otherwise. Let $x_{m,l}^{(n)} \in \{0,1\}$ be 1 if beam $n$ of BS $l$ is assigned to UAV $m$. The desired signal power received at UAV $m$ from its serving BS $l$ using beam $n$:
\begin{equation}
\label{eq:signal_power}
S_{m,l,n} = P_{\mathrm{tx}} |h_{m,l,n}|^2 g_{m,l,n},
\end{equation}
where $P_{\mathrm{tx}}$ is the transmit power allocated to each active beam, $|h_{m,l,n}|^2$ is the channel gain obtained from the CT, and $g_{m,l,n} = 10^{G_{5\mathrm{G}}(\theta_m, \phi_m)/10}$ represents the linear antenna gain of beam $n$ from BS $l$ toward UAV $m$.

\noindent The interference experienced by UAV $m$ is the sum of powers from all other transmissions. This includes inter-cell interference from other BSs and intra-cell interference from other beams of the same serving BS. The aggregated interference can be modeled as:
\begin{equation}
\label{eq:interference_model}
I_{m} = \sum_{l'=1}^{L} \sum_{\substack{m'=1 \\ m' \neq m}}^{M} \sum_{n'=1}^{N} \beta_{m',l'} x_{m',l'}^{(n')} P_{\mathrm{tx}} |h_{m,l',n'}|^2 g_{m,l',n'}.
\end{equation}

\noindent The downlink Signal-to-Interference-plus-Noise Ratio (SINR) for UAV $m$ associated with BS $l$ and beam $n$ can be modeled as:
\begin{equation}
\label{eq:sinr_model}
\mathrm{SINR}_{m} = \frac{\beta_{m,l} x_{m,l}^{(n)} S_{m,l,n}}{I_m + \sigma^2},
\end{equation}
where $\sigma^2 = N_0 B$ is the thermal noise power over bandwidth $B$. Finally, the achievable throughput for UAV $m$ is given by the Shannon-Hartley theorem:
\begin{equation}
\label{eq:throughput_model}
    R_m = B \log_2 \left( 1 + \mathrm{SINR}_{m} \right).
\end{equation}

\noindent We adopt the following practical assumptions for system deployment. a) First, we assume that drone positions are perfectly integrated within the CT. Existing literature shows such position feedback can be directly obtained via drone mounted location sensors \cite{charan2025sensing}. Moreover, 5G also supports network assisted positioning through dedicated position reference signals \cite{dwivedi2021positioning}. b). Also, adopting coordinated architectures similar to those used in vehicular millimeter wave (mmWave) systems \cite{tarafder2023deep}, we consider a centralized radio resource management (RRM) system, where all the BSs are connected to a centralized edge controller, that hosts CT and DRL and reliable feedback link among BS and controller exists to exchange control information. c) Moreover reliable control feedback link between BS and drone exists which will be used to inform drone about its serving BS/beam index \cite{sun2025aerial}. d) Semi-persistent scheduling is considered where assigned BSs and beams to the UAVs remain unchanged on the duration of seconds to avoid frequent switching operations.
\section{Optimization Problem Formulation and Proposed RRM Framework}
\label{sec:ppo_framework}
\subsection{Problem Formulation}
In this paper, our objective is to optimize the UAV-BS-beam association policy that maximizes the aggregated network throughput, $R_{\text{sum}} = \sum_{m=1}^{M} R_m$. This problem is formulated as the following combinatorial optimization problem:
\vspace{-1em}
\begin{small}
\begin{alignat}{2}
(\mathcal{P}_0) \; \max_{\substack{\{\beta_{m,l}, \\ x_{m,l}^{(n)}\}}} & \sum_{m=1}^{M} R_m \label{eq:opt_obj}\\
\text{s.t.} \quad
&\text(C1): \sum_{l=1}^{L} \beta_{m,l} = 1, \forall m, 
&& \hspace{-0.2cm}\text(C2): \sum_{m=1}^{M} \beta_{m,l} \leq N, \forall l \label{eq:c2_opt_obj}\\
& \hspace{-0.25cm}\text(C3): \sum_{n=1}^{N} x_{m,l}^{(n)} = \beta_{m,l}, \forall m,l,
&& \text(C4):\sum_{m=1}^{M} x_{m,l}^{(n)} \leq 1, \forall l,n. 
\end{alignat}
\end{small}
\vspace{-0.1em}
In $(\mathcal{P}_0)$, constraint C1 ensures each UAV associates with one BS. Constraint C2 limits the number of UAVs associated with each BS to at most $N$, since each active beam can serve at most one UAV. This constraint implicitly enforces UAV load balancing among the active BSs. Constraints C3 and C4 jointly guarantee exclusive, single-beam assignment at each BS. Solving $(\mathcal{P}_0)$ using conventional optimization techniques entails two fundamental challenges. \textbf{Challenge 1:} $(\mathcal{P}_0)$ is equivalent to a three-dimensional matching problem, which is NP-hard. Specifically, obtaining the optimal solution requires exhaustive search, whose computational complexity grows exponentially with the number of UAVs. As a result, optimally solving $(\mathcal{P}_0)$ is computationally infeasible for real-time operation in dynamic aerial corridor environments.
\textbf{Challenge 2:} Solving $(\mathcal{P}_0)$ optimally requires accurate global CSI $\{h_{m,l,n}\}$, across all UAVs, BSs, and beams. In practice, acquiring such global CSI is highly non-trivial. Standard cellular systems rely on two primary CSI acquisition mechanisms: (i) uplink sounding reference signals (SRSs), exploiting time-division duplexing (TDD) and uplink-downlink reciprocity, and (ii) beam sweeping combined with measurement reporting. In the considered scenario, where a UAV may maintain strong LOS links with multiple BSs, SRS-based CSI acquisition leads to excessive pilot overhead, severe pilot contamination, increased computational complexity, and stringent inter-BS synchronization requirements. Moreover, due to the high mobility of UAVs, uplink CSI obtained via SRS may become outdated and unreliable for downlink transmission, thereby violating the uplink–downlink reciprocity assumption. Similarly, beam sweeping leads to significant signaling overhead, as each UAV must measure the received signal strength of multiple beams from multiple BSs and report these measurements via uplink control channels. Therefore, even if an optimal algorithm for solving $(\mathcal{P}_0)$ were available, the requirement for accurate global CSI constitutes a major bottleneck to the practical deployment of such optimization-based solutions.

To address these challenges and solve P0 efficiently and in a scalable manner, we adopt a CT-guided learning approach. Specifically, to tackle \textbf{Challenge 1}, we propose a two-stage solution framework that (i) identifies a feasible set of candidate beams (beam codebook vectors) at each BS to support the UAVs, and (ii) selects the active beam set per BS that satisfies the association constraints while maximizing the network sum capacity. To address \textbf{Challenge 2}, we exploit a CT\footnote{We assume that, prior to deployment, the CT is well calibrated using field measurement data, enabling accurate generation of SS multipath CSI. The CT calibration process itself is beyond the scope of this work.} to acquire global CSI using low-fidelity feedback, such as the three-dimensional positions of UAVs and BSs. Further details of two stage optimization are given as follows:

\noindent \textbf{Stage 1: Beam Gain Optimization:}
For each BS–UAV–beam combination, we determine the
optimal horizontal scan angle $\phi_{\text{scan}}$ that maximizes the 3GPP antenna gain such that it focuses its maximum energy to UAV's physical location. In this first stage, the objective is to determine, for each beam in the link between a 5G BS and a UAV, the scan angle $\phi_{\text{scan}}$ that maximizes the gain defined in Eq.~\eqref{eq:total_gain} by solving the optimization problem: 
\begin{equation}\label{eq:phi_scan}
\phi_{\text{scan}}^{\star} = \arg\max_{\phi_{\text{scan}} \in [-\pi,\pi]}
G_{5\mathrm{G}}(\theta, \phi_{\text{scan}}).
\end{equation}
This non-convex problem is solved using a widely known stochastic optimization algorithm \emph{dual annealing},
which alternates between global exploration (simulated annealing transitions that escape local maxima) and local refinement
(deterministic search in promising regions). The optimized beam gains for every UAV–BS–beam triplet are then used in Stage 2 via Eq.~\eqref{eq:signal_power}. The operation of dual annealing algorithm per link $(m,l,n)$ is described in Algorithm~\ref{alg:simulated_annealing}.

\begin{algorithm}[t]
\small 
\caption{Dual Annealing with Local Refinement for $\varphi_{\mathrm{scan}}$}
\label{alg:simulated_annealing}
\begin{algorithmic}[1]
\State \textbf{Initialization:} Sample $\varphi_{\mathrm{scan}}^{(0)}\!\sim\!\mathcal{U}[-\pi,\pi]$ and set initial temperature $T_0$.
\State Evaluate initial objective value $G_{\mathrm{5G}}(\theta,\varphi_{\mathrm{scan}}^{(0)})$.
\For{$t=0$ to $t_{\max}$}
    \State \textbf{Global (SA) Step:} Propose $\tilde{\varphi}=\varphi_{\mathrm{scan}}^{(t)}+\Delta$,
    \State where $\Delta$ is a symmetric perturbation (e.g., Cauchy or Gaussian).
    \State Wrap $\tilde{\varphi}$ to $[-\pi,\pi]$.
    \State Compute $\Delta g = G_{\mathrm{5G}}(\theta,\tilde{\varphi}) - G_{\mathrm{5G}}(\theta,\varphi_{\mathrm{scan}}^{(t)})$.
    \State Set acceptance probability $p = \min\{1, \exp(\Delta g/T_t)\}$.
    \If{Uniform random number $\le p$}
        \State Accept new state: $\varphi_{\mathrm{scan}}^{(t+1)} \gets \tilde{\varphi}$.
    \Else
        \State Keep current state: $\varphi_{\mathrm{scan}}^{(t+1)} \gets \varphi_{\mathrm{scan}}^{(t)}$.
    \EndIf
    \State \textbf{Cooling:} Update $T_{t+1}=\alpha T_t$ with $\alpha\!\in\!(0,1)$.
    \If{$T_t$ is small or improvement stagnates}
        \State \textbf{Local Refinement:} Invoke bounded local optimizer (e.g., golden-section or Brent) near the incumbent $\varphi_{\mathrm{scan}}^{(t)}$.
    \EndIf
\EndFor
\State \textbf{Stopping and Caching:} Stop when $T_t$ or improvement $<\varepsilon$; store $G^{\star}_{i,m,l}=G_{\mathrm{5G}}(\theta,\varphi_{\mathrm{scan}}^{\star})$ and $\varphi_{\mathrm{scan}}^{\star}$.
\end{algorithmic}
\end{algorithm}

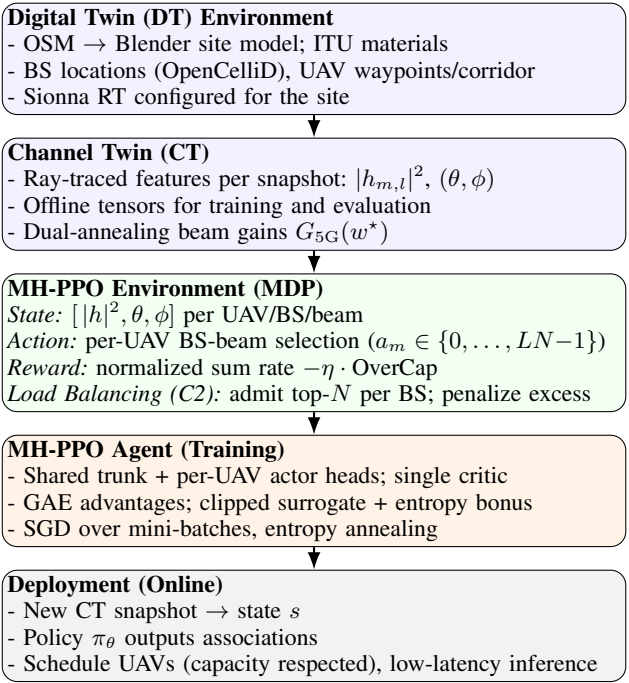
\begin{figure}[t]
\centering
\begin{tikzpicture}[
  >=Latex, node distance=8mm, font=\small,
  box/.style={draw, rounded corners=2mm, align=left, inner sep=2pt,
              minimum height=10mm, text width=0.92\columnwidth},
  off/.style={fill=blue!5}, env/.style={fill=green!5},
  agent/.style={fill=orange!10}, deploy/.style={fill=gray!10},
  line/.style={-Latex, thick}
]
\node[box, off] (dt) {\textbf{Digital Twin (DT) Environment}\\
- OSM $\rightarrow$ Blender site model; ITU materials\\
- BS locations (OpenCelliD), UAV waypoints/corridor\\
- Sionna RT configured for the site};

\node[box, off, below=3mm of dt] (ct) {\textbf{Channel Twin (CT)}\\
- Ray-traced features per snapshot: $|h_{m,l}|^2$, $(\theta,\phi)$\\
- Offline tensors for training and evaluation\\
- Dual-annealing beam gains $G_{\mathrm{5G}}(w^\star)$};

\node[box, env, below=3mm of ct] (env) {\textbf{MH-PPO Environment (MDP)}\\
\emph{State:} $[\,|h|^2,\theta,\phi]$ per UAV/BS/beam\\
\emph{Action:} per-UAV BS-beam selection ($a_m\in\{0,\ldots,LN{-}1\}$)\\
\emph{Reward:} normalized sum rate $-\eta\cdot\text{OverCap}$\\
\emph{Load Balancing (C2):} admit top-$N$ per BS; penalize excess};

\node[box, agent, below=3mm of env] (ppo) {\textbf{MH-PPO Agent (Training)}\\
- Shared trunk + per-UAV actor heads; single critic\\
- GAE advantages; clipped surrogate + entropy bonus\\
- SGD over mini-batches, entropy annealing};

\node[box, deploy, below=3mm of ppo] (dep) {\textbf{Deployment (Online)}\\
- New CT snapshot $\rightarrow$ state $s$\\
- Policy $\pi_\theta$ outputs associations\\
- Schedule UAVs (capacity respected), low-latency inference};

\draw[line] (dt) -- (ct);
\draw[line] (ct) -- (env);
\draw[line] (env) -- (ppo);
\draw[line] (ppo) -- (dep);
\end{tikzpicture}
\caption{Workflow of the proposed framework.}
\label{fig:workflow}
\end{figure}

\noindent \textbf{Stage 2: UAV–BS–Beam Association via PPO:} From the feasible set of candidate beams per BS, where each beam corresponds to a potential UAV-BS connection identified in Stage 1, Stage 2 selects the final subset of UAV-BS-beam associations that (i) satisfy the constraints of $\mathcal{P}_0$ and (ii) maximize the network sum capacity. Owing to the mutual interference and interdependence among the selected links, this combinatorial problem cannot be reduced to a simple bipartite assignment formulation. Consequently, we cast the UAV-BS-beam association problem as a learning task and seek to learn an optimal association policy using an RL approach. A trained MH-PPO agent accepts the precomputed beam gains obtained from Stage 1, CT-derived channel features, AoAs, and outputs per-UAV BS-beam assignments. In this stage, we enforce the constraints i.e., load balancing and beam exclusivity introduced in ($\mathcal{P}_0$) within the UAV DRL environment through capacity penalties. This enables low-latency, real-time inference while retaining near-optimal throughput performance. Our proposed two-stage design combines the physics-consistent beamforming gains from dual annealing (Stage~1) with a scalable, learning-based association policy from MH-PPO (Stage~2). Together, they provide a DT-driven optimization pipeline that is both accurate and real-time capable, overcoming the limitations of purely optimization-based methods. Fig~\ref{fig:workflow} illustrates the detailed workflow developed to efficiently solve problem $(\mathcal{P}_0)$. Next, we introduce the fundamentals of PPO-based RL, followed by the agent design and algorithm development. 

\subsection{RL and PPO Fundamentals}
Reinforcement learning (RL) formalizes the problem of sequential decision-making as an interaction between an agent and a stochastic environment, described mathematically as a Markov Decision Process (MDP) $M = \langle \mathcal{S}, \mathcal{A}, P, r, \gamma \rangle$~\cite{SuttonBarto2018}. At each discrete time $t$, the agent observes a state $s_t \in \mathcal{S}$, selects an action $a_t \in \mathcal{A}$ according to its policy $\pi$, receives a scalar reward $r_t = r(s_t,a_t)$, and transitions to the next state $s_{t+1} \sim P(\cdot|s_t,a_t)$. The objective is to learn a policy $\pi_\theta$, parameterized by weights $\theta$, that maximizes the expected discounted return:
\begin{equation}
J(\pi_\theta) = \mathbb{E}_{\tau \sim \pi_\theta}\left[\sum_{t=0}^{\infty} \gamma^t r(s_t,a_t)\right],
\end{equation}
where $\gamma \in [0,1]$ is the discount factor balancing immediate and future rewards. Unlike value-based methods (e.g., DQN), which learn value functions and derive the policy, PPO directly parameterizes the policy as $\pi_{\theta}(a_{t}|s_{t})$, with parameters $\theta$. Policy performance is often measured by the state-value function $V^\pi(s) = \mathbb{E}^\pi[\sum_{t=0}^\infty \gamma^t r(s_t,a_t)\mid s_0=s]$ and the action-value function $Q^\pi(s,a) = \mathbb{E}^\pi[\sum_{t=0}^\infty \gamma^t r(s_t,a_t)\mid s_0=s, a_0=a]$. Once we have the state-value and action-value functions, an advantage function denoted by $A^\pi(s,a) = Q^\pi(s,a) - V^\pi(s)$ quantifies the benefit of taking action $a$ relative to the policy’s average. Key properties of the advantage function include:
\begin{itemize}
  \item If \(A^\pi(s,a) > 0\), the action \(a\) is better than average and the policy should increase its probability.
  \item If \(A^\pi(s,a) < 0\), the action \(a\) is worse than average and the policy should decrease its probability.
  \item If \(A^\pi(s,a) = 0\), the action \(a\) is average and the policy does not update its probability for this action.
\end{itemize}
These properties form the intuition behind the policy gradient update, which for differentiable policies, is formalized by the policy gradient theorem and provides a principled direction for optimizing the policy parameters as follows:
\begin{equation}
\nabla_\theta J(\pi_\theta) = \mathbb{E}_{s\sim d^\pi, a\sim \pi_\theta}\left[ \nabla_\theta \log \pi_\theta(a|s)\, A^\pi(s,a) \right].
\end{equation}
The reasons behind choosing PPO \cite{schulman2017ppo} to solve the UAV-BS-beam association policy learning problem are as follows. First,  the agent must select a BS and a beam direction from a finite set at each decision step, resulting in a discrete action space comprised of candidate BS-beam pairs. PPO can naturally handle discrete action spaces by parameterizing its policy as a categorical distribution, allowing effective exploration and robust learning in this context. Furthermore, PPO is an ON-policy algorithm that use fresh trajectories for policy updates, yielding stable performance in dynamic DT environments where channel conditions evolve rapidly.
By contrast, Off-policy methods (e.g., DQN, TD3, SAC) utilizes past experience to improve sample efficiency\footnote{While recent off-policy methods such as SAC and TD3 offer improved stability over DQN through techniques like twin critics and entropy regularization, we focus our empirical comparison on the foundational on-policy (PPO) versus off-policy (DQN) paradigms to isolate the impact of policy update mechanisms. Comprehensive comparison with advanced off-policy methods remains a direction for future work.}. However, these algorithms can suffer from distribution mismatch, where the experience data generated by an earlier policy diverges from the current policy, potentially degrading the performance in fast-changing UAV networks.

Moreover, exploration is another core requirement for an effective RL algorithm. Sufficient environment exploration for the DRL agent is essential to circumvent premature convergence to suboptimal BS-beam associations. PPO facilitates exploration explicitly by incorporating an entropy regularization, which extends the probability mass over candidate BS-beam pairs during early training. This entropy-driven approach is particularly effective when actions are categorical, whereas common off-policy methods often rely on strategies such as Gaussian or parameter-space noise (or $\epsilon$-greedy in DQN) for exploration, which may be less suitable in such settings. Another critical issue is the bias-variance trade-off in advantage estimation. Advantage functions can be estimated using either Monte Carlo returns or temporal-difference (TD) methods. Monte Carlo returns provide unbiased estimates but have high variance, while TD methods reduce variance at the cost of increased bias. To address this trade-off, we employ Generalized Advantage Estimation (GAE)~\cite{schulman2016gae}, which interpolates between low-variance, biased TD-errors and high-variance, unbiased Monte Carlo returns via a parameter $\Lambda \in [0,1]$:
\begin{align}
\delta_t &= r_t + \gamma V_\phi(s_{t+1}) - V_\phi(s_t), \\
\hat{A}_t^{\text{GAE}(\gamma,\Lambda)} &= \sum_{l=0}^\infty (\gamma\Lambda)^l \delta_{t+l},
\end{align}
where $V_\phi$ denotes the learned critic network. By tuning $\Lambda$, we can balance bias and variance, where lower $\Lambda$ values favor lower variance but higher bias, while higher $\Lambda$ values favor unbiased but high variance Monte Carlo returns. Beyond advantage estimation, PPO prevents destabilizing large policy updates through a clipped surrogate objective that constrains the policy change at each iteration:
\begin{equation}
L^{\text{CLIP}}(\theta) = \mathbb{E}_t \!\left[\min\!\left(\rho_t(\theta)\hat{A}_t, \;\text{clip}(\rho_t(\theta), 1-\epsilon, 1+\epsilon)\hat{A}_t\right)\right],
\end{equation}
where $\rho_t(\theta)=\pi_\theta(a_t|s_t)/\pi_{\theta_{\text{old}}}(a_t|s_t)$ represents the probability ratio between the current and previous policies, and $\epsilon$ is a clipping hyperparameter. This clipped objective provides the stability benefits of trust region methods while maintaining a simpler first-order implementation. To further ensure sufficient exploration in the vast BS-beam action space, an entropy bonus is incorporated into the loss function, encouraging the policy to maintain stochasticity and avoid premature convergence to suboptimal associations.

\begin{figure*}[t]
\centering
\includegraphics[width=1.5\columnwidth]{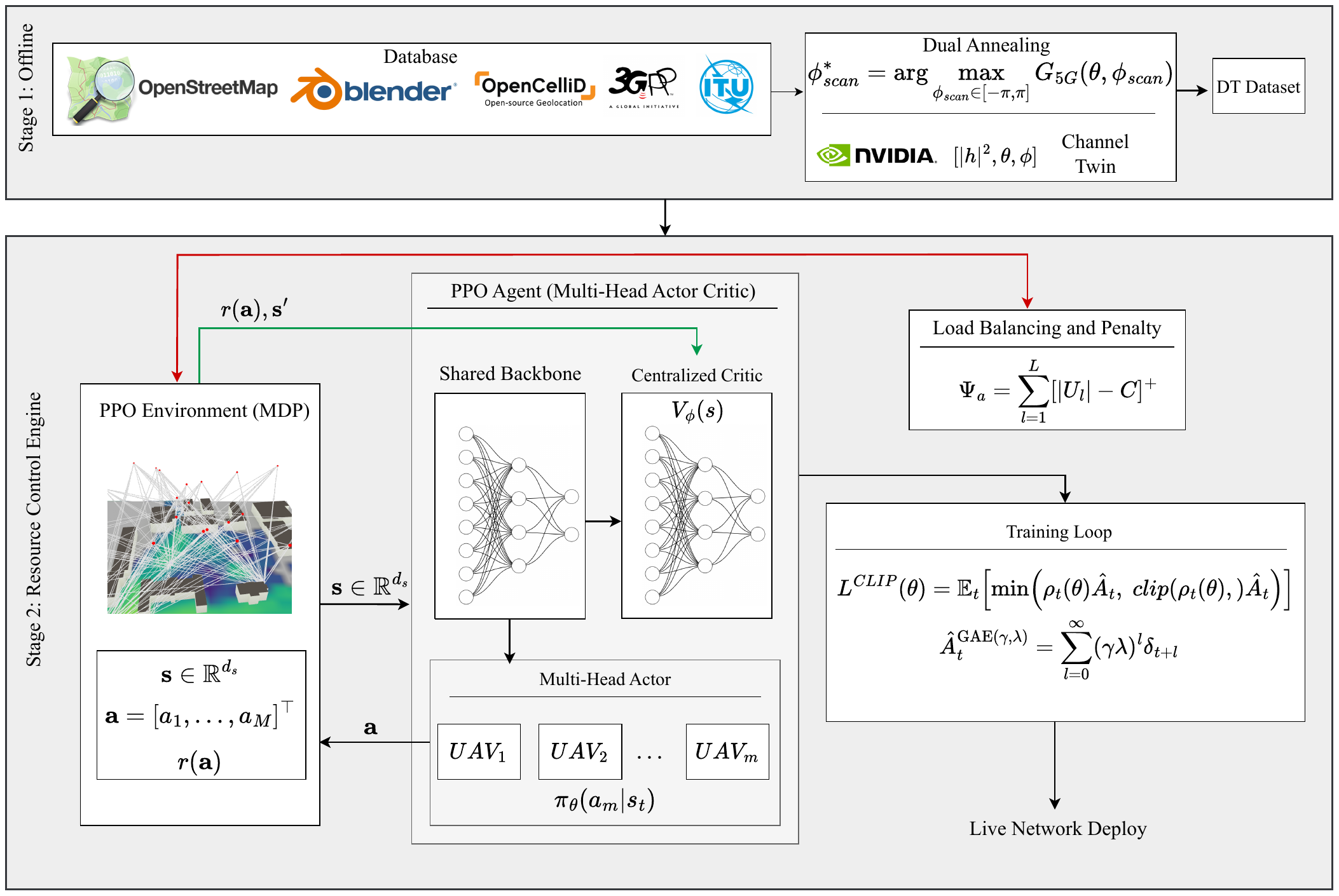}
\caption{Proposed Multi-Head PPO based UAV-BS-Beam Association Architecture}
\label{fig:System Architecture}
\end{figure*}

\subsection{MH-PPO Agent Architecture}
Our proposed MH-PPO agent employs a multi-head actor-critic architecture that establishes a balance between decentralized execution and centralized learning. The network comprises three principal components: a shared feature extraction backbone, $M$ decentralized actor heads, and a centralized critic network.

\subsubsection{Shared Feature Extraction Backbone}

The shared trunk processes the high-dimensional state vector $\mathbf{s} \in \mathbb{R}^{d_s}$, where $d_s = 3MLN$ corresponds to the flattened CT-derived features (channel gain, azimuth AoA, and zenith AoA for each UAV-BS-beam triple). The backbone consists of a sequence of fully connected layers with Rectified Linear Unit (ReLU) activations:
\begin{equation}
    \mathbf{z} = f_{\text{base}}(\mathbf{s}; \boldsymbol{\theta}_{\text{base}}),
\end{equation}
where the transformation is defined recursively as:
\begin{align}
    \mathbf{h}_1 &= \sigma(\mathbf{W}_1 \mathbf{s} + \mathbf{b}_1), \\
    \mathbf{h}_2 &= \sigma(\mathbf{W}_2 \mathbf{h}_1 + \mathbf{b}_2), \\
    \mathbf{h}_3 &= \sigma(\mathbf{W}_3 \mathbf{h}_2 + \mathbf{b}_3), \\
    \mathbf{z} &= \sigma(\mathbf{W}_4 \mathbf{h}_3 + \mathbf{b}_4),
\end{align}
with $\sigma(\cdot)$ denoting the ReLU activation function. The weight matrices define progressive dimensionality reduction: $\mathbf{W}_1 \in \mathbb{R}^{1024 \times d_s}$, $\mathbf{W}_2 \in \mathbb{R}^{512 \times 1024}$, $\mathbf{W}_3 \in \mathbb{R}^{256 \times 512}$, and $\mathbf{W}_4 \in \mathbb{R}^{128 \times 256}$. Note that weight matrix dimensions follow the standard convention $\mathbf{W} \in \mathbb{R}^{d_{\text{out}} \times d_{\text{in}}}$, where the output dimension precedes the input dimension. The resulting latent representation $\mathbf{z} \in \mathbb{R}^{d_z}$ with $d_z = 128$ encodes hierarchical features relevant to the association task.

\subsubsection{Decentralized Actor Heads}

The shared latent representation $\mathbf{z}$ feeds into $M$ independent actor heads, one designated for each UAV $m \in \mathcal{M}$. Each actor head $f_{\text{actor}}^{(m)}$ implements a linear transformation that maps $\mathbf{z}$ to logits over the $LN$-dimensional BS-beam action space:
\begin{equation}
    \boldsymbol{\ell}_m = \mathbf{W}_{\text{actor}}^{(m)} \mathbf{z} + \mathbf{b}_{\text{actor}}^{(m)}, \quad \mathbf{W}_{\text{actor}}^{(m)} \in \mathbb{R}^{LN \times d_z},
\end{equation}
where $\boldsymbol{\ell}_m \in \mathbb{R}^{LN}$ represents the unnormalized log-probabilities for UAV $m$. A subsequent softmax operation produces the categorical probability distribution over actions:
\begin{equation}
    \pi_{\theta}(a_m | \mathbf{s}) = \frac{\exp(\ell_{m,a_m})}{\sum_{k=0}^{LN-1} \exp(\ell_{m,k})}, \quad a_m \in \{0, \ldots, LN-1\}.
\end{equation}
This decentralized structure enables parallel, independent action selection across all UAVs while leveraging the shared learned state representation.

\subsubsection{Centralized Critic Network}

The critic network estimates the state-value function $V_\phi(\mathbf{s})$, which approximates the expected cumulative reward from state $\mathbf{s}$ under the current policy. Unlike a simple linear projection, we employ a multi-layer perceptron (MLP) to provide sufficient capacity for accurate value estimation in complex environments:
\begin{equation}
    V_\phi(\mathbf{s}) = f_{\text{critic}}(\mathbf{z}; \boldsymbol{\phi}),
\end{equation}
where the critic transformation is given by:
\begin{align}
    \mathbf{v}_1 &= \sigma(\mathbf{W}_v^{(1)} \mathbf{z} + \mathbf{b}_v^{(1)}), \\
    \mathbf{v}_2 &= \sigma(\mathbf{W}_v^{(2)} \mathbf{v}_1 + \mathbf{b}_v^{(2)}), \\
    V_\phi(\mathbf{s}) &= \mathbf{W}_v^{(3)} \mathbf{v}_2 + b_v^{(3)},
\end{align}
with $\mathbf{W}_v^{(1)} \in \mathbb{R}^{64 \times d_z}$, $\mathbf{W}_v^{(2)} \in \mathbb{R}^{32 \times 64}$, and $\mathbf{W}_v^{(3)} \in \mathbb{R}^{1 \times 32}$. The deeper critic architecture reduces bias in advantage estimation, particularly for high-dimensional state spaces characteristic of multi-UAV networks.

\subsubsection{Computational Complexity Analysis}

The proposed architecture effectively decomposes the joint action space into per-UAV decisions. For a na\"ive joint-action formulation, the action space cardinality scales exponentially as $\mathcal{O}((LN)^M)$, rendering direct policy parameterization intractable for large $M$. Our factorized design reduces this to $\mathcal{O}(M \cdot LN)$, as each actor head independently samples from its local categorical distribution. This linear scaling facilitates deployment in dense UAV scenarios without architectural modifications, since additional UAVs require only the instantiation of corresponding actor heads operating on the common latent representation $\mathbf{z}$.

The complete parameter set comprises the shared backbone parameters $\boldsymbol{\theta}_{\text{base}}$, the per-UAV actor parameters $\{\boldsymbol{\theta}_{\text{actor}}^{(m)}\}_{m=1}^{M}$, and the critic parameters $\boldsymbol{\phi}$. During training, all parameters are updated jointly via gradient descent on the PPO objective. At inference, only the actor components are required, enabling low-latency BS-beam association decisions suitable for real-time 6G operations.

\begin{algorithm}[t]
\caption{DT-Assisted UAV-BS-Beam Association with MH-PPO}
\label{alg:MH-PPO}
\begin{algorithmic}[1]
\Require Digital Twin (DT) scene, BSs $\mathcal{L}$, UAVs $\mathcal{M}$, beam codebooks $\{\mathcal{C}_l\}$, PPO hyperparameters $(\gamma, \Lambda, \epsilon, \beta, \alpha_\theta, \alpha_\phi, K, B)$.
\Ensure Trained PPO policy $\pi_\theta$ for UAV-BS-beam association.

\Statex \textbf{I. Channel Twin Construction (Offline)}
\For{each scenario $s = 1, \ldots, S$}
    \State Generate SS 3D map of UAV corridor.
    \State Place BSs and UAVs; run Sionna RT to obtain $\{|h|^2, \theta, \phi\}$.
\EndFor

\Statex \textbf{II. Beam-Gain Optimization}
\For{each BS $l$ and beam index $n$}
    \State Apply dual annealing to solve $\phi^\star_{\text{scan}} = \arg\max_{\phi_{\text{scan}}} G_{\text{5G}}(\theta, \phi_{\text{scan}})$.
    \State Store optimized beam gains $G_{\text{5G}}(w^\star_{n,l}; \theta, \phi)$.
\EndFor

\Statex \textbf{III. MH-PPO Training}
\State Initialize policy $\pi_\theta$, value function $V_\phi$.
\Repeat
    \State Sample scenario; observe state $\mathbf{s} = [\text{vec}(|\mathbf{H}|^2), \text{vec}(\boldsymbol{\Phi}), \text{vec}(\boldsymbol{\Theta})]^\top$.
    \State Sample joint action $\mathbf{a} \sim \pi_\theta(\cdot|\mathbf{s})$ and store $\log \pi_\theta(\mathbf{a}|\mathbf{s})$.
    \State Apply load balancing: admit top-$N$ UAVs per BS by received power.
    \State Compute reward $r(\mathbf{a}) = \frac{1}{M}\sum_{m \in \mathcal{A}} R_m - \eta \cdot \Psi_\mathbf{a}$.
    \State Store transition $(\mathbf{s}, \mathbf{a}, r, \mathbf{s}')$ in batch $\mathcal{D}$.
    \For{each timestep $t$ in $\mathcal{D}$}
        \State Compute TD error: $\delta_t = r_t + \gamma \hat{V}_{t+1} - \hat{V}_t$.
        \State Estimate advantage via GAE: $\hat{A}_t = \sum_{l=0}^{\infty}(\gamma\Lambda)^l \delta_{t+l}$.
        \State Compute target return: $\hat{R}_t = \hat{A}_t + \hat{V}_t$.
    \EndFor
    \State Normalize advantages $\hat{A}_t$ to zero mean and unit variance.
    \State Set $\theta_{\text{old}} \gets \theta$.
    \For{epoch $k = 1$ to $K$}
        \State Partition $\mathcal{D}$ into mini-batches of size $B$.
        \For{each mini-batch $\mathcal{B}$}
            \State Compute importance ratio: $\rho_t(\theta) = \exp[\log \pi_\theta(a_t|s_t) - \log \pi_{\theta_{\text{old}}}(a_t|s_t)]$.
            \State Compute clipped surrogate loss:
            \Statex \quad\quad $L^{\text{CLIP}} = \frac{1}{|\mathcal{B}|}\sum_{t \in \mathcal{B}} \min\left(\rho_t \hat{A}_t, \text{clip}(\rho_t, 1-\epsilon, 1+\epsilon)\hat{A}_t\right)$.
            \State Compute critic loss: $L^V = \frac{1}{|\mathcal{B}|}\sum_{t \in \mathcal{B}} (V_\phi(s_t) - \hat{R}_t)^2$.
            \State Compute entropy bonus: $L^{\text{ENT}} = \frac{1}{|\mathcal{B}|}\sum_{t \in \mathcal{B}} \mathcal{H}(\pi_\theta(\cdot|s_t))$.
            \State Update actor: $\theta \gets \theta + \alpha_\theta \nabla_\theta (L^{\text{CLIP}} + \beta L^{\text{ENT}})$.
            \State Update critic: $\phi \gets \phi - \alpha_\phi \nabla_\phi L^V$.
        \EndFor
    \EndFor
    \State Optionally anneal entropy coefficient $\beta$.
\Until{training budget exhausted or convergence reached}

\Statex \textbf{IV. Deployment}
\State For new DT snapshot, extract state $\mathbf{s}$ with CT-derived features.
\State Evaluate $\pi_\theta$ to obtain UAV-BS-beam associations.
\State Schedule UAVs respecting load-balancing constraints.
\end{algorithmic}
\end{algorithm}
\vspace{-1.0em}
\subsection{RL Environment Design and Algorithm Development}
We formulate the UAV-BS-beam association as a single-step MDP, where the environment processes each joint action and computes a reward signal that balances throughput maximization against capacity constraints.

\noindent\textbf{State and Action Spaces:} The state $\mathbf{s} \in \mathbb{R}^{d_s}$ aggregates CT-derived channel features across all UAV-BS-beam combinations. For each triple $(m, l, n)$ corresponding to UAV $m \in \mathcal{M}$, BS $l \in \mathcal{L}$, and beam $n \in \mathcal{N}$, the state comprises three features: the channel gain $|h_{m,l,n}|^2$, mean azimuth angle of arrival $\phi_{m,l,n}$, and mean zenith angle of arrival $\theta_{m,l,n}$. These tensors are flattened and concatenated to form the input vector:
\begin{equation}
    \mathbf{s} = \left[ \mathrm{vec}(|\mathbf{H}|^2)^\top, \mathrm{vec}(\boldsymbol{\Phi})^\top, \mathrm{vec}(\boldsymbol{\Theta})^\top \right]^\top \in \mathbb{R}^{3MLN},
\end{equation}
where $\mathrm{vec}(\cdot)$ denotes the vectorization operator, $|\mathbf{H}|^2 \in \mathbb{R}^{M \times L \times N}$ is the channel gain tensor, and $\boldsymbol{\Phi}, \boldsymbol{\Theta} \in \mathbb{R}^{M \times L \times N}$ are the azimuth and zenith AoA tensors, respectively. The optimized beam gains $G_{5\mathrm{G}}$, precomputed via dual annealing in Stage~1, are stored separately and accessed during reward computation rather than included in the state representation\footnote{While actions influence rewards but not states, spatio-temporal CSI correlations preserve the Markov property, validating PPO as a robust optimizer for this DT framework.}.


The agent produces a joint action $\mathbf{a} = [a_1, \ldots, a_M]^\top$, where each per-UAV action $a_m \in \{0, \ldots, LN-1\}$ indexes a unique BS-beam pair. Each action $a_m$ admits a unique factorization into BS and beam indices via:
\begin{equation}
    l_m = \lfloor a_m / N \rfloor, \quad n_m = a_m \mod N,
\end{equation}
where $l_m \in \{0, \ldots, L-1\}$ and $n_m \in \{0, \ldots, N-1\}$ denote the selected BS and beam indices for UAV $m$, respectively. This encoding collapses the two-dimensional decision $(l_m, n_m)$ into a single categorical choice, enabling efficient sampling from the actor head's probability distribution.

\noindent\textbf{Capacity-Constrained Admission Control:}
Upon receiving the joint action $a$, the environment enforces constraint ~\eqref{eq:c2_opt_obj} through a power-based admission policy. For each BS $l \in \{1,\dots,L\}$ we define the contender set $\mathcal{U}_l = \{\,m \in \{1,\dots,M\} \mid l_m = l\,\}$ as the subset of UAVs requesting service from BS $l$. When the cardinality $|\mathcal{U}_l|$ exceeds the per-BS capacity $C$, the environment prioritizes UAVs by their instantaneous received signal strength. Specifically, for each contender $m \in \mathcal{U}_l$, the desired received power is computed as:
\begin{equation}
\label{eq:desired_power}
p_m = P_{\mathrm{tx}} g_{m,l_m,n_m} 10^{\frac{b_{m,l_m,n_m}}{10}},
\end{equation}
where $P_{\mathrm{tx}}$ denotes the transmit power in watts, $g_{m,l,n} \in \mathbb{R}_+$ represents the linear-scale channel gain (encompassing path loss and shadowing) between UAV $m$ and BS $l$ on beam $n$, and $b_{m,l,n}$ is the corresponding beamforming gain. The per-BS admitted set is then determined by selecting the top-$C$ UAVs with maximum received power as follows:
\begin{equation}
\mathcal{A}_l =
\begin{cases}
\mathcal{U}_l, & \text{if } |\mathcal{U}_l| \le C, \\
\arg\max\limits_{\substack{\mathcal{S}\subseteq \mathcal{U}_l \\ |\mathcal{S}|= C}} \sum\limits_{m\in \mathcal{S}} p_m, & \text{otherwise}.
\end{cases}
\end{equation}
UAV admission decisions are made independently at each BS. Hence, the sets $\{\mathcal{A}_l\}_{l=1}^L$ are mutually disjoint due to the one-to-one UAV–BS association constraint. The global active set therefore comprises of:
\begin{equation}
\mathcal{A} = \bigcup_{l=1}^L \mathcal{A}_l \subseteq \{1,\dots,M\}.
\end{equation}
UAVs in the complement set $\mathcal{A}^c = \{1,\dots,M\} \setminus \mathcal{A}$ are denied service and achieve zero throughput. The environment then quantifies constraint violations by summing excess UAVs across all BSs by:
\vspace{-1.0em}
\begin{equation}
\label{eq:overcap}
\Psi_{a} = \sum_{l=1}^{L} \left[ |\mathcal{U}_l| - C \right]^+ ,
\end{equation}

\noindent where $[\cdot]^+ = \max(0, \cdot)$ denotes the positive-part operator.

\noindent\textbf{Reward Function Design:} The scalar reward signal $r: \mathcal{A} \to \mathbb{R}$ balances network throughput against constraint satisfaction. For each admitted UAV $m \in \mathcal{A}$, the instantaneous data rate $R_m$ is evaluated via the Shannon capacity formula under co-channel interference, as specified in Eq.~\eqref{eq:throughput_model}. The reward function is constructed as:
\begin{equation}
    r(\mathbf{a}) = \frac{1}{M} \sum_{m \in \mathcal{A}} R_m - \eta \Psi_{\mathbf{a}},
    \label{eq:reward}
\end{equation}
where the first term represents the average per-UAV throughput normalized by the total number of UAVs $M$, and the second term penalizes capacity constraint violations. The penalty coefficient $\eta = 10^3$ regularizes the trade-off between capacity enforcement and throughput maximization, while the overcapacity count $\Psi_{\mathbf{a}}$ is defined in Eq.~\eqref{eq:overcap}.

This reward structure steers gradient-based policy optimization toward high-utilization, load-balanced topologies, whereas infeasible associations receive penalties proportional to the number of rejected UAVs, while feasible associations satisfying $|\mathcal{U}_l| \leq C$ for all $l$ are rewarded according to their aggregate rate. The normalization by $M$ rather than $|\mathcal{A}|$ implicitly penalizes solutions that leave many UAVs unserved, encouraging the agent to maximize both coverage and throughput. 

The proposed UAV–BS–beam association procedure is summarized in Algorithm~\ref{alg:MH-PPO}, which comprises both offline and online phases. In the offline phase, Algorithm~\ref{alg:MH-PPO} constructs the CT and generates the training dataset (lines 1–3), followed by training the MH-PPO agent using GAE (lines 7–30). In the online phase, Algorithm~\ref{alg:MH-PPO} first determines the optimal beam gains by executing the Stage I optimization (lines 4–6). It then selects a near-optimal UAV–BS–beam association by deploying the trained PPO agent (lines 31–33).

\begin{table}[t]
\centering
\caption{Simulation Parameters and MH-PPO Hyperparameters}
\label{tab:sim_params}
\begin{tabular}{l l}
\hline
\textbf{Parameter} & \textbf{Value} \\
\hline
\multicolumn{2}{c}{\textit{System Parameters}} \\
\hline
Simulation area & $330 \times 300$ m (Howard University) \\
Carrier Frequency & 3.5 GHz \\
UAV altitude range $\xi$ & 60-100 m \\
Number of UAVs $M$ & [20, 25, 30]\\
Number of BSs $L$ & 4 \\
Antennas per BS & $N=16$ (UPA, 3GPP TR~37.840) \\
BS transmit power $P$ & 40 W (46 dBm)\\
System bandwidth $W$ & 20 MHz \\
Noise power density & $-174$ dBm/Hz \\
Channel modeling & NVIDIA Sionna RT \\
Beam optimization & Dual annealing  \\
Reward penalty $\eta$ & $1\times 10^3$ \\
Antenna pattern & 3GPP Sectorized \\
Vertical tilt angle & $15^\circ$ \\
Max element gain ($G_{E,\max}$) & -8 dBi \\
Elevation beamwidth ($\theta_{3\text{dB}}$) & $65^\circ$ \\
Azimuth beamwidth ($\phi_{3\text{dB}}$) & $90^\circ$ \\
Front-to-back ratio ($A_m$) & 30 dB \\
Side-lobe level limit ($SL_{A,V}$) & 30 dB \\
\hline
\multicolumn{2}{c}{\textit{MH-PPO Hyperparameters}} \\
\hline
Discount factor $\gamma$ & 0.99 \\
GAE parameter $\Lambda$ & 0.97 \\
Clipping parameter $\epsilon$ & 0.15 \\
Entropy coefficient $\beta$ & 0.2 $\rightarrow$ 0.005 (annealed) \\
Learning rate (actor/critic) & $3\times 10^{-4}$ \\
Optimizer & Adam \\
Batch size $B$ & 2056 \\
Epochs per update $K$ & 12 \\
Training timesteps & $35\times 10^6$ \\
Neural net (shared base) & [1024, 512, 256, 128] \\
Actor heads & One per UAV ($L\times N = 64$ outputs) \\
Critic head & [128, 64, 32, 1] \\
\hline
\end{tabular}
\end{table}
\vspace{-1.0em}
\section{Performance Evaluation}
\label{sec:results}
In this section, we evaluate the performance of the proposed DT-enabled MH-PPO framework. We benchmark our agent against classical optimization/heuristic methods and a learning-based baseline. The evaluation is structured to analyze not only the raw throughput performance, but also the agent's ability to generalize to entirely unseen network scenarios. A summary of key simulation parameters are provided in Table~\ref{tab:sim_params}.
\vspace{-20pt}
\subsection{Baseline Schemes and Evaluation Protocol}
In this paper, we benchmark our proposed MH-PPO-based association framework by comparing it with five baseline schemes: Vanilla Deep Q Network (DQN), Two-Stage Hungarian Algorithm, Max-Gain, Closest-BS, Random Assignment. Note that all baselines are evaluated under identical CT-derived channel realizations, ensuring a fair comparison of algorithmic performance rather than channel model fidelity.
\vspace{-1.0em}
\subsection{Learning-based baseline}
\noindent\textit{\textbf{Baseline 1 (Vanilla DQN)}}: We employ a factorized multi-head DQN agent trained on the same CT-generated dataset as the proposed MH-PPO framework as a baseline off-policy deep reinforcement learning algorithm. The DQN architecture uses a shared feature extraction backbone followed by $M$ independent Q-value heads, one for each UAV, is structurally identical to the proposed MH-PPO agent. Actions are selected greedily as follows, and each head outputs Q-values over the $LN$-dimensional BS-beam action space:
\begin{equation}
    a_m = \arg\max_{k \in \{0, \ldots, LN-1\}} Q_m(\mathbf{s}, k; \boldsymbol{\omega}),
\end{equation}
where $Q_m(\mathbf{s}, k; \boldsymbol{\omega})$ denotes the Q-value for UAV $m$ selecting action $k$ in state $\mathbf{s}$, parameterized by weights $\boldsymbol{\omega}$. The DQN agent is trained using experience replay with a buffer capacity of $3 \times 10^5$ transitions and a minimum of $10^4$ samples before training begins. Target network parameters are synchronized every $10^3$ steps via hard updates. Exploration follows a cosine-annealed $\epsilon$-greedy schedule, where $\epsilon$ decays from $1.0$ to $0.01$ over $5 \times 10^5$ steps according to:
\begin{equation}
    \epsilon(t) = \epsilon_{\mathrm{end}} + \frac{\epsilon_{\mathrm{start}} - \epsilon_{\mathrm{end}}}{2} \left(1 + \cos\left(\frac{\pi \cdot \min(t, T_{\mathrm{decay}})}{T_{\mathrm{decay}}}\right)\right),
\end{equation}
where $T_{\mathrm{decay}} = 5 \times 10^5$. This baseline evaluates whether off-policy methods can match the stability and generalization of on-policy PPO in the high-dimensional UAV association problem.
\vspace{-1.0em}
\subsection{Optimization-Based Baseline}
\noindent\textit{\textbf{Baseline 2 (Two-Stage Hungarian Algorithm)}}: This method represents our prior work~\cite{tarafder2025digital} and serves as a near-optimal combinatorial optimization benchmark.
\vspace{-1.0em}
\subsection{Heuristic Baselines}
\noindent\textit{\textbf{Baseline 3 (Maximum Channel Gain)}}: Instead of using instantaneous beam-level measurements, this heuristic prioritizes BS selection based on average channel quality. We first determine the BS index that offers highest average channel gain among all beams for each UAV $m$ as follows: 
\begin{equation}
    l_m^* = \arg\max_{l \in \mathcal{L}} \frac{1}{N} \sum_{n=1}^{N} |h_{m,l,n}|^2.
\end{equation}
Afterwards, the algorithm selects the beam with the highest effective gain $\mathcal{G}_{{m,l_m^*,n}}$ among those not yet assigned to other UAVs once the preferred BS has been identified. The algorithm switches to the next-best BS in the ranked list if all beams at the preferred BS are already occupied. UAVs that are taken into consideration earlier have access to a greater variety of beams because they are processed in a sequential manner. Although channel-aware decision-making is incorporated into this greedy approach, the spatial distribution of UAVs and the ensuing inter-cell interference patterns are not taken into account.

\noindent\textit{\textbf{Baseline 4 (Closest-BS)}}: 
Baseline 4 corresponds to a geometry based heuristic that associates UAVs with BSs solely based on spatial proximity. Specifically, each UAV $m$ is initially assigned and associated with the BS that minimizes the Euclidean distance as follows:
\begin{equation}
    l_m^* = \arg\min_{l \in \mathcal{L}} \|\mathbf{p}_m^{\mathrm{UAV}} - \mathbf{p}_l^{\mathrm{BS}}\|_2,
\end{equation}
where $\mathbf{p}_m^{\mathrm{UAV}}, \mathbf{p}_l^{\mathrm{BS}} \in \mathbb{R}^3$ denote the coordinates of UAV $m$ and BS $l$, respectively. Once the serving BS is selected, the beam which yields the maximum effective gain $\mathcal{G}_{{m,l_m^*,n}}$ among the unassigned candidates are selected. If the nearest BS has no available beams, the algorithm proceeds to the next closest alternative. Although this heuristic incurs low computational overhead and intuitively associates each UAV with its nearest BS which is optimal in clutter-free environments, its performance degrades in dense or cluttered urban scenarios because it does not account for SS CSI.

\noindent\textit{\textbf{Baseline 5 (Random Assignment)}}: We also include a simple heuristic that assigns each UAV to a randomly and uniformly chosen BS-beam pair as a lower bound for reference:
\vspace{-0.4em}
\begin{equation} a_m \sim \mathrm{Uniform}\{0, 1, \ldots, LN-1\}, \quad \forall m \in \mathcal{M}. \end{equation}

This naive baseline does not leverage any form of geometric relationships, nor beam gains, nor channel state information. Most importantly, it does not impose any form of beam exclusivity, hence multiple UAVs can select the very same BS-beam pairs and may account for conflicts which leads to poor performance. Although random assignment is not deployable, it provides a performance floor useful to quantify the benefit brought in by intelligent resource allocation. Table~\ref{tab:baseline_summary} summarizes the computational complexity of each methods.

\begin{figure}[t]
\centering
\includegraphics[width=0.8\columnwidth]{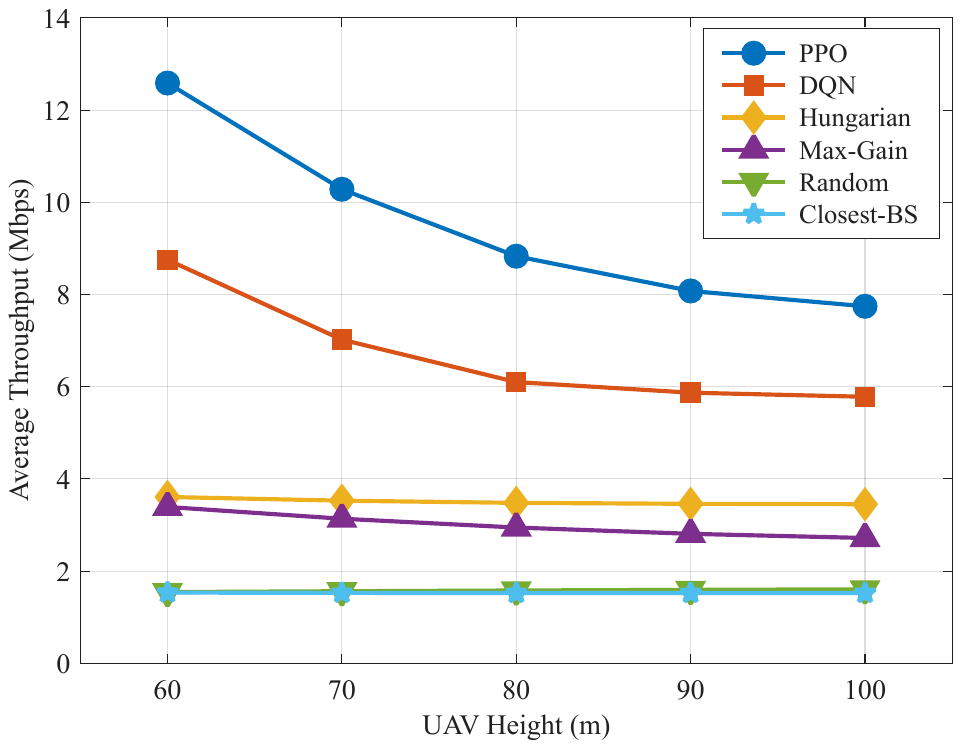}
\caption{Altitude vs. per-UAV throughput, $M$ = 20}
\label{fig:throughput_20uavs}
\end{figure}

\begin{figure}[t]
\centering
\includegraphics[width=0.8\columnwidth]{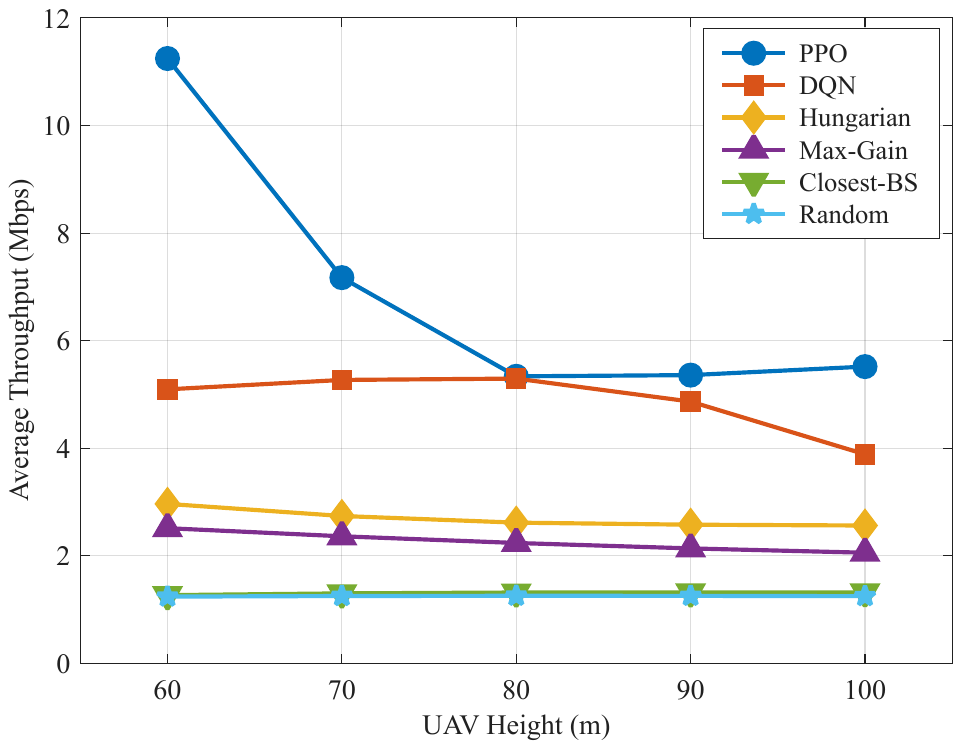}
\caption{Altitude vs. per-UAV throughput, $M$ = 25}
\label{fig:throughput_25uavs}
\end{figure}

\begin{figure}[t!]
\centering
\includegraphics[width=0.8\columnwidth]{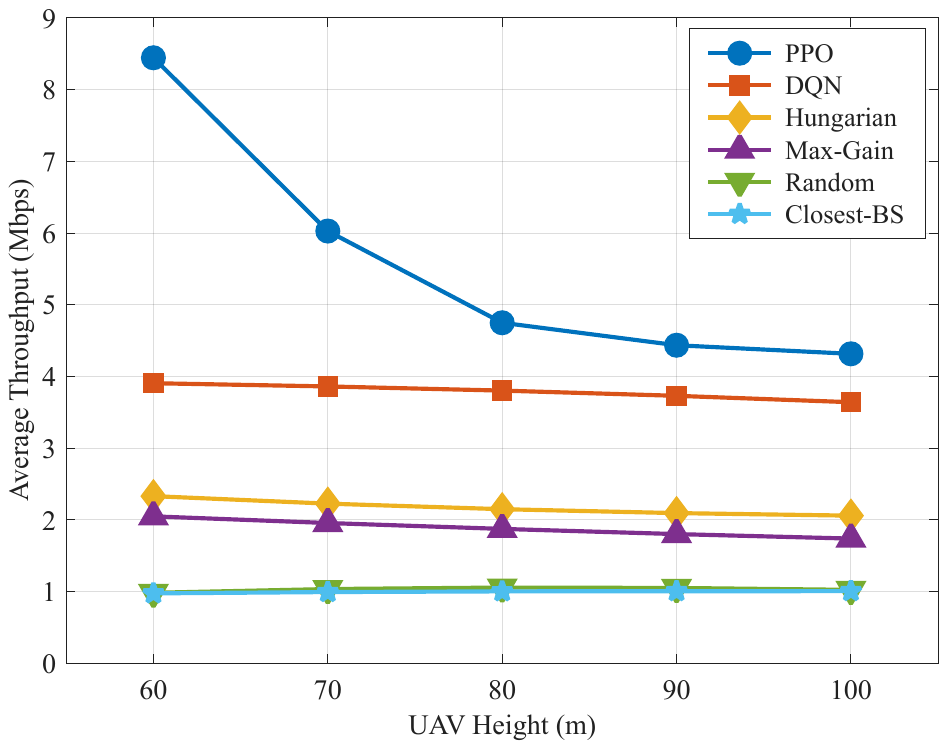}
\caption{Altitude vs. per-UAV throughput, $M$ = 30}
\label{fig:throughput_30uavs}
\end{figure}
\vspace{-1.0em}
\subsection{Experimental Setup}
\begin{figure*}[t]
\centering
\begin{subfigure}[b]{0.32\textwidth}  
    \includegraphics[width=\textwidth]{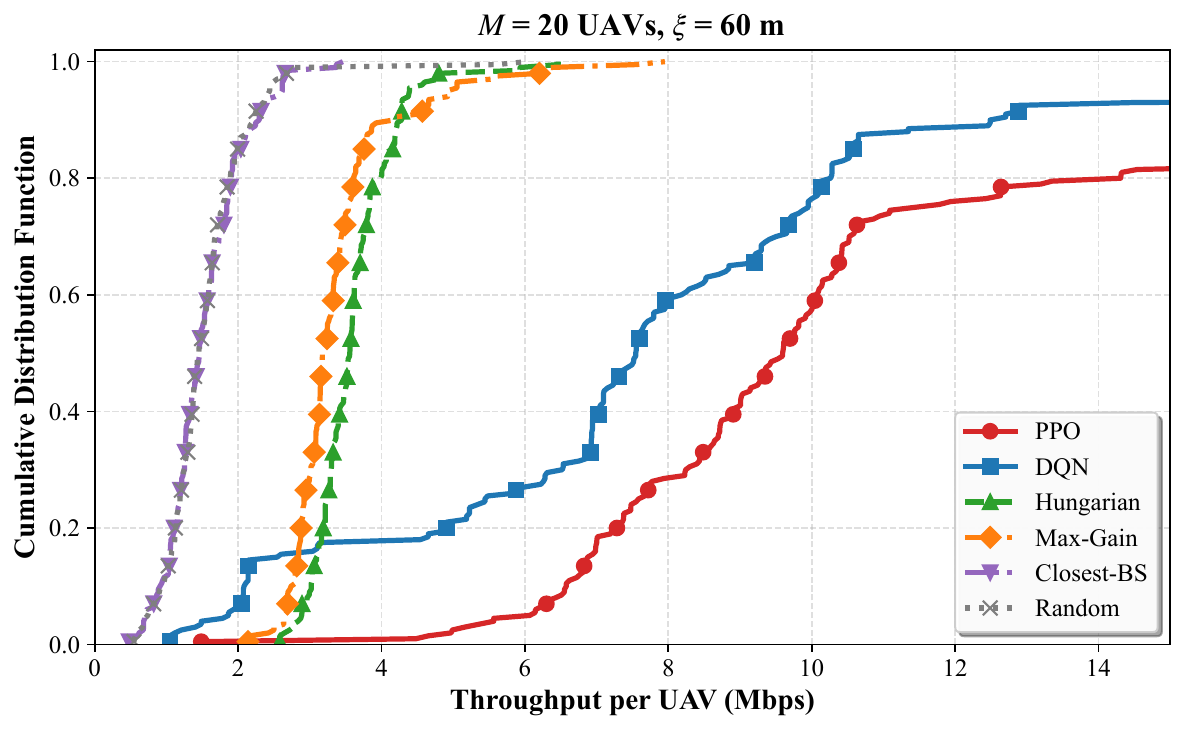}
    \caption{20 UAVs, 60m}
    \label{fig:sub1}
\end{subfigure}
\hfill
\begin{subfigure}[b]{0.32\textwidth}
    \includegraphics[width=\textwidth]{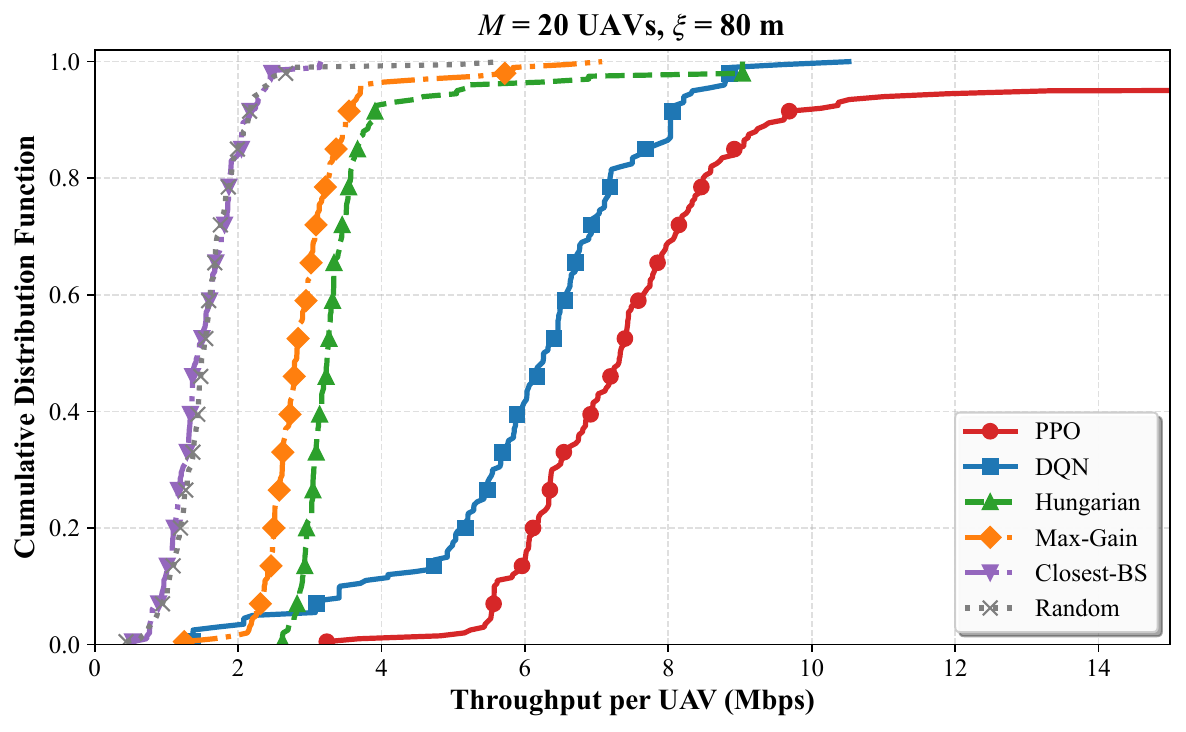}
    \caption{20 UAVs, 80m}
    \label{fig:sub2}
\end{subfigure}
\hfill
\begin{subfigure}[b]{0.32\textwidth}
    \includegraphics[width=\textwidth]{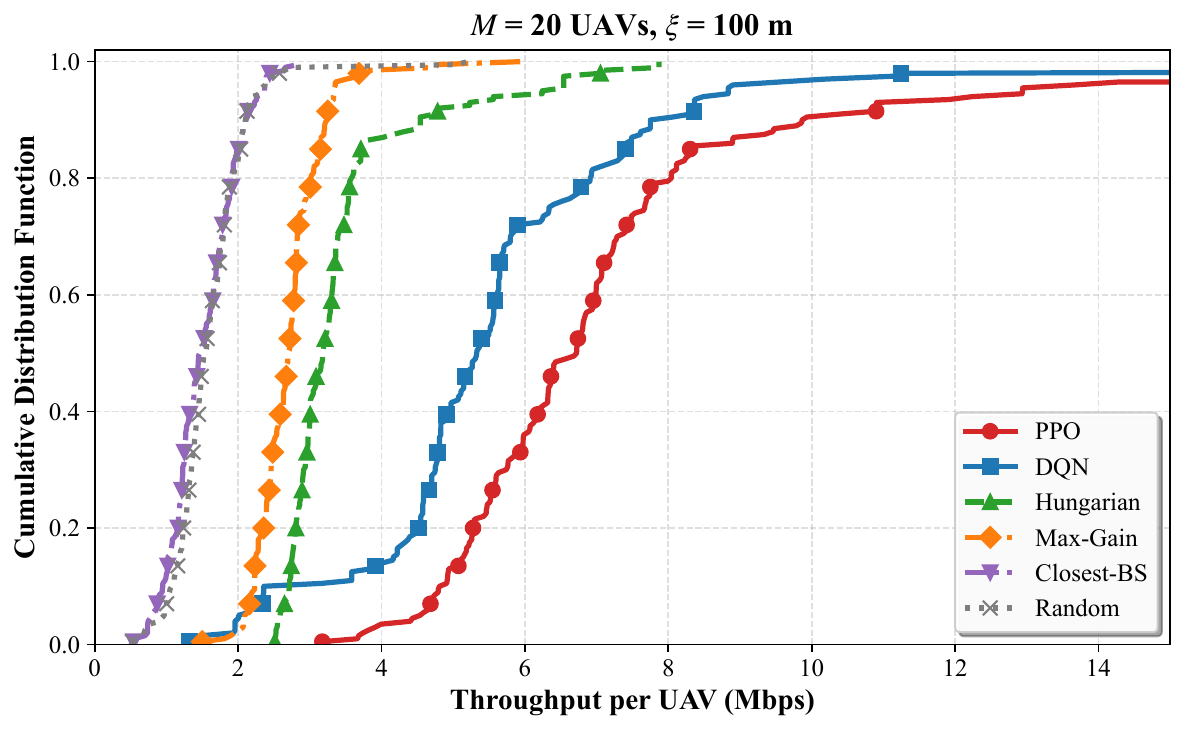}
    \caption{20 UAVs, 100m}
    \label{fig:sub3}
\end{subfigure}

\vspace{0.3cm}

\begin{subfigure}[b]{0.32\textwidth}
    \includegraphics[width=\textwidth]{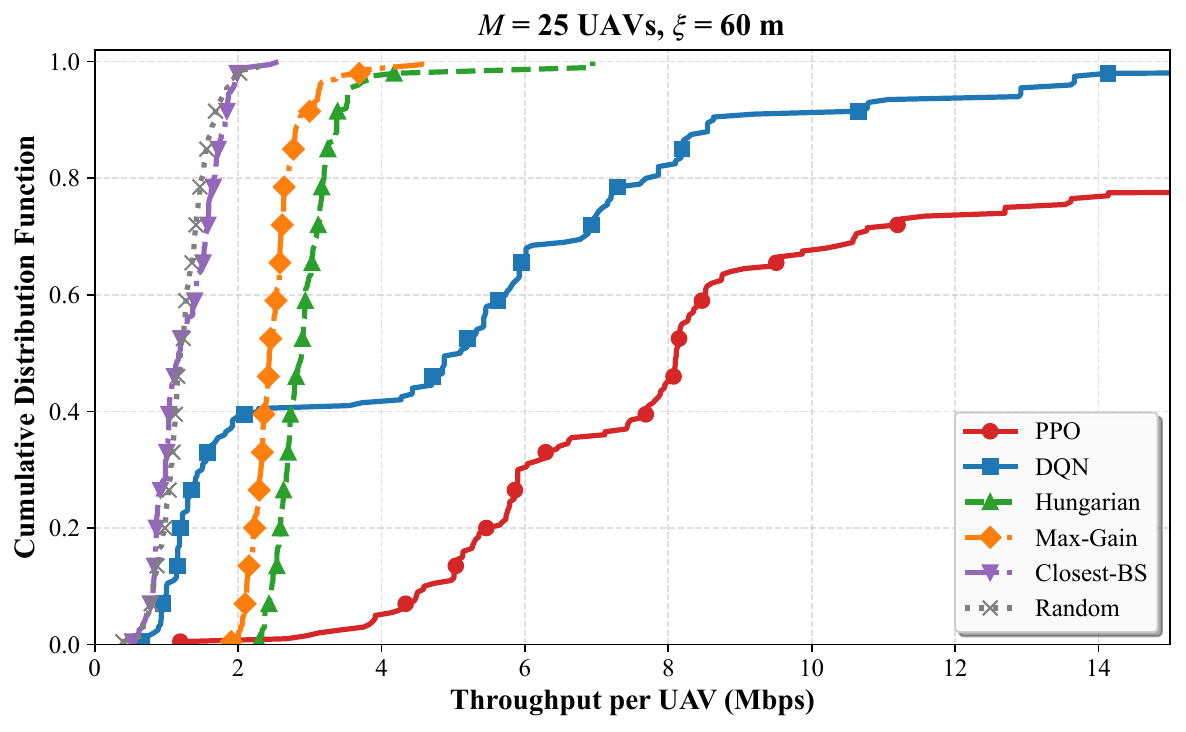}
    \caption{25 UAVs, 60m}
    \label{fig:sub4}
\end{subfigure}
\hfill
\begin{subfigure}[b]{0.32\textwidth}
    \includegraphics[width=\textwidth]{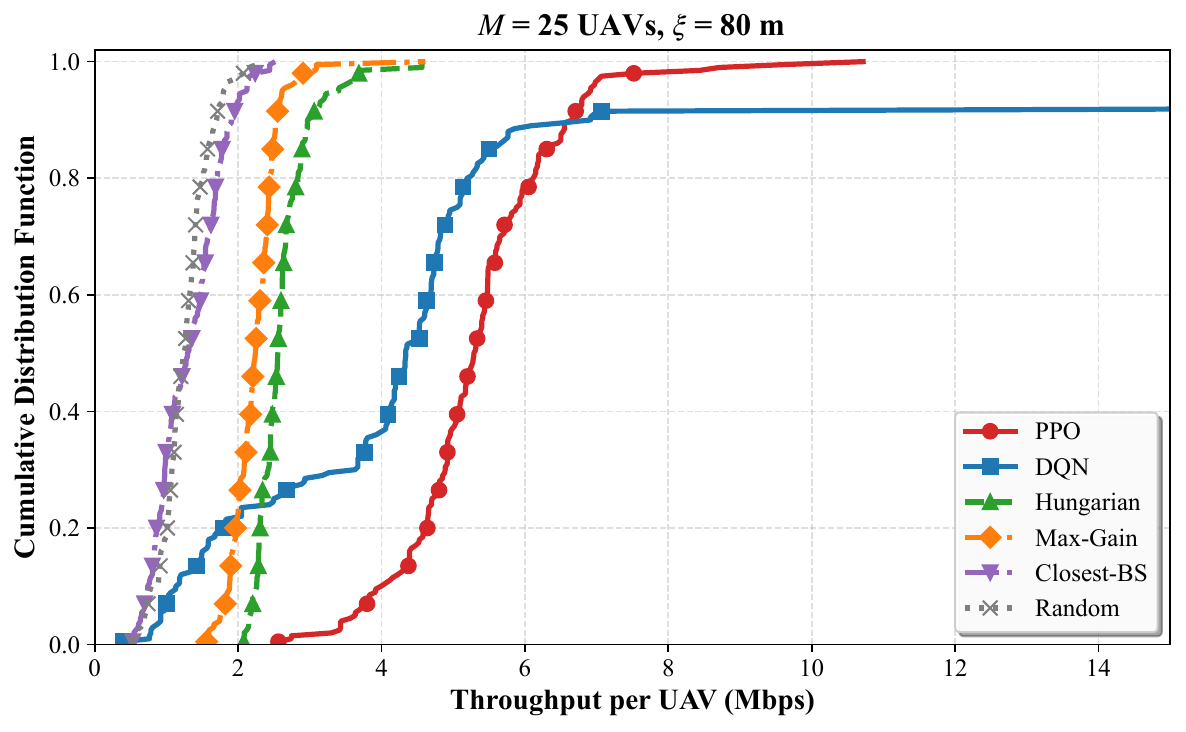}
    \caption{25 UAVs, 80m}
    \label{fig:sub5}
\end{subfigure}
\hfill
\begin{subfigure}[b]{0.32\textwidth}
    \includegraphics[width=\textwidth]{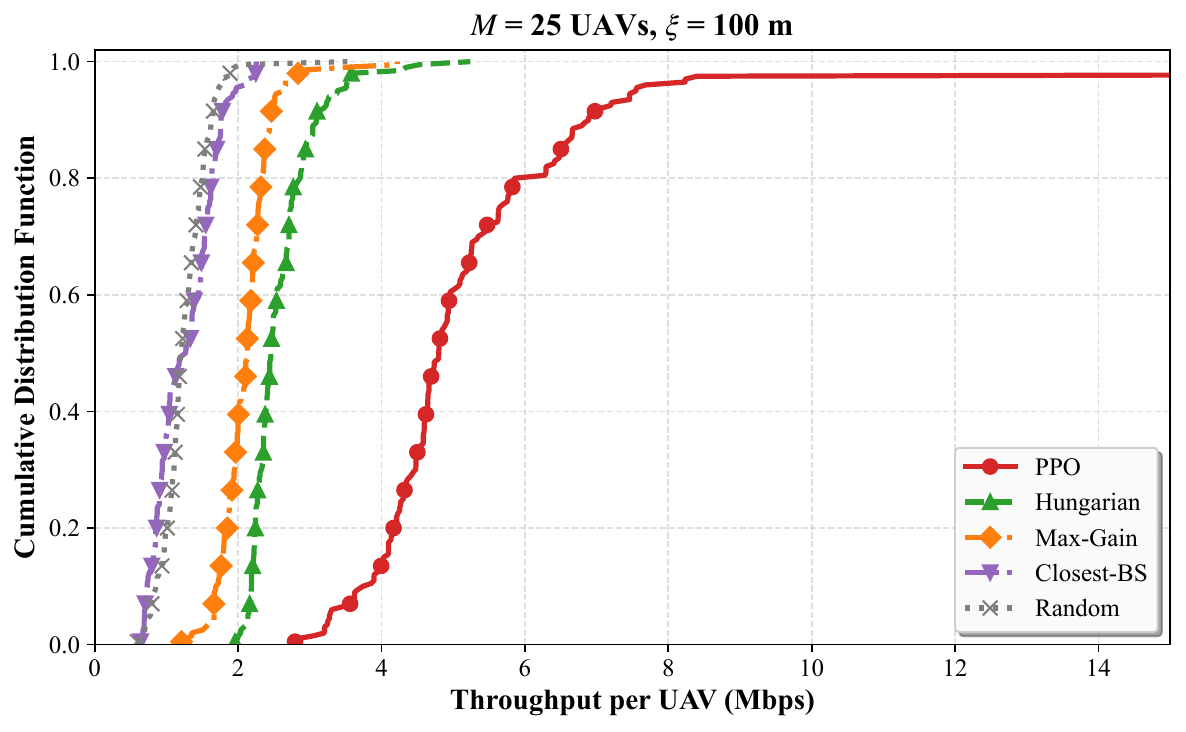}
    \caption{25 UAVs, 100m}
    \label{fig:sub6}
\end{subfigure}

\vspace{0.3cm}

\begin{subfigure}[b]{0.32\textwidth}
    \includegraphics[width=\textwidth]{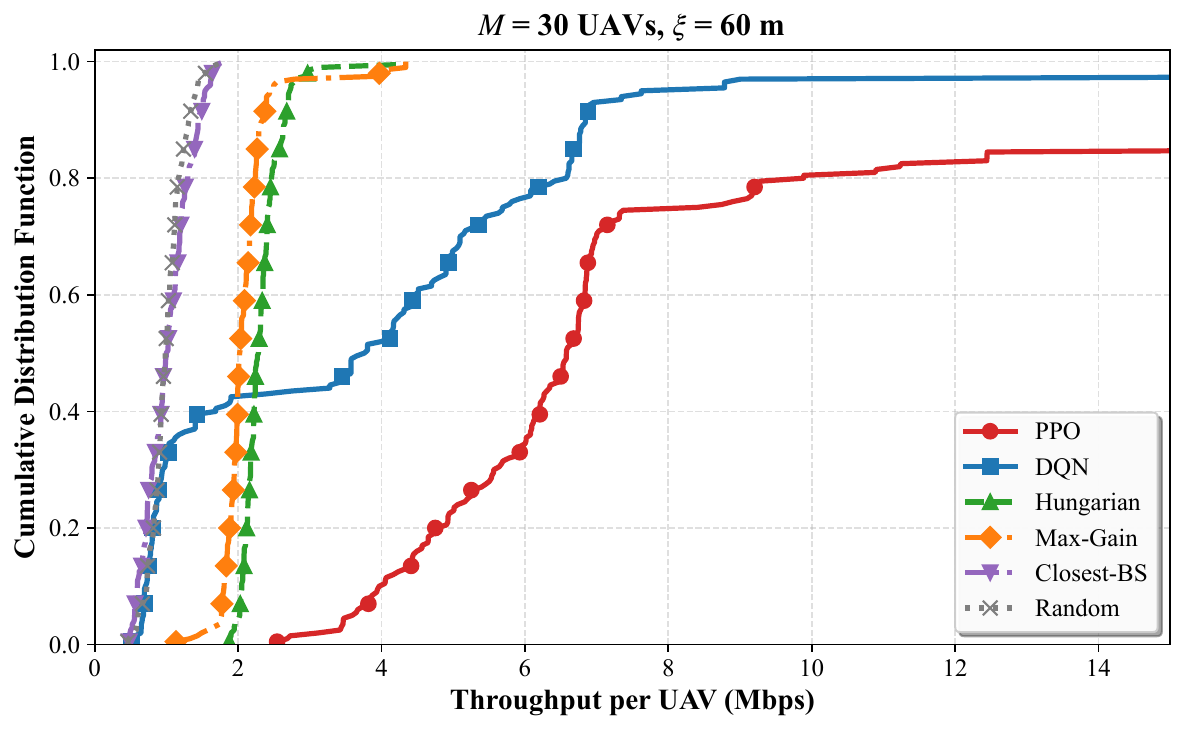}
    \caption{30 UAVs, 60m}
    \label{fig:sub7}
\end{subfigure}
\hfill
\begin{subfigure}[b]{0.32\textwidth}
    \includegraphics[width=\textwidth]{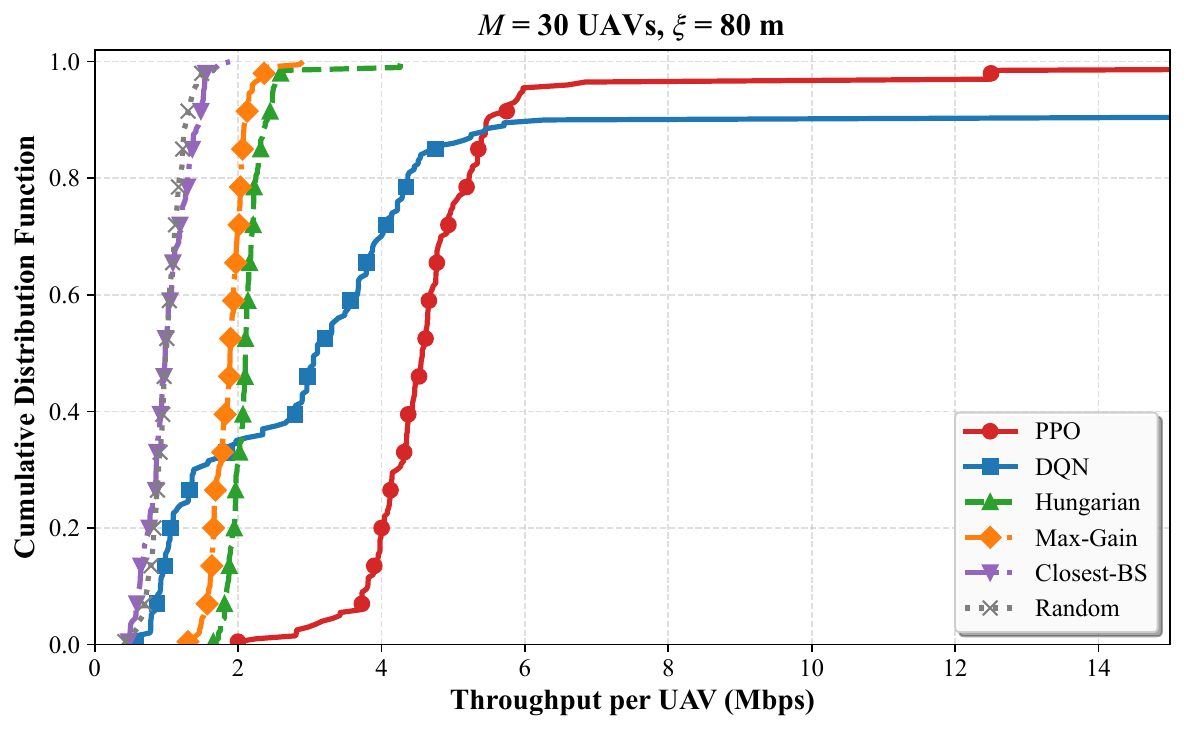}
    \caption{30 UAVs, 80m}
    \label{fig:sub8}
\end{subfigure}
\hfill
\begin{subfigure}[b]{0.32\textwidth}
    \includegraphics[width=\textwidth]{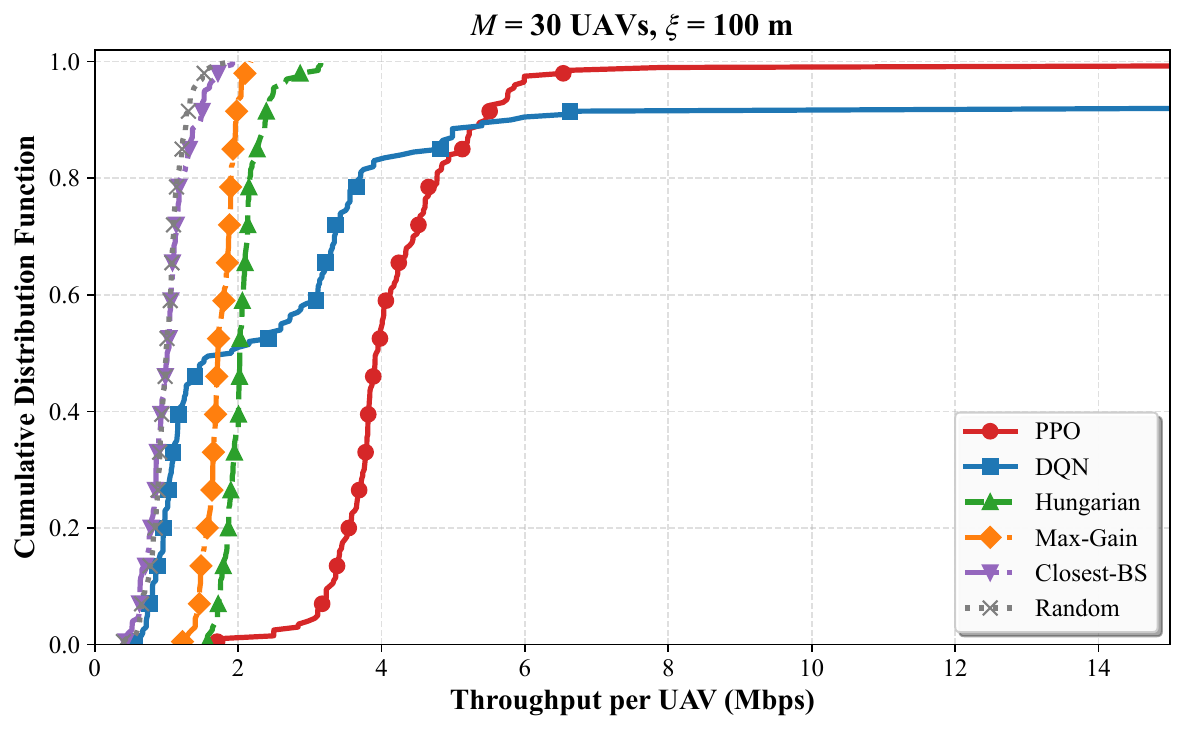}
    \caption{30 UAVs, 100m}
    \label{fig:sub9}
\end{subfigure}

\caption{CDF of throughput performance across different UAV counts and heights.}
\label{fig:cdf_subplots}
\end{figure*}

\begin{figure}[t]
\centering
\includegraphics[width=0.9\columnwidth]{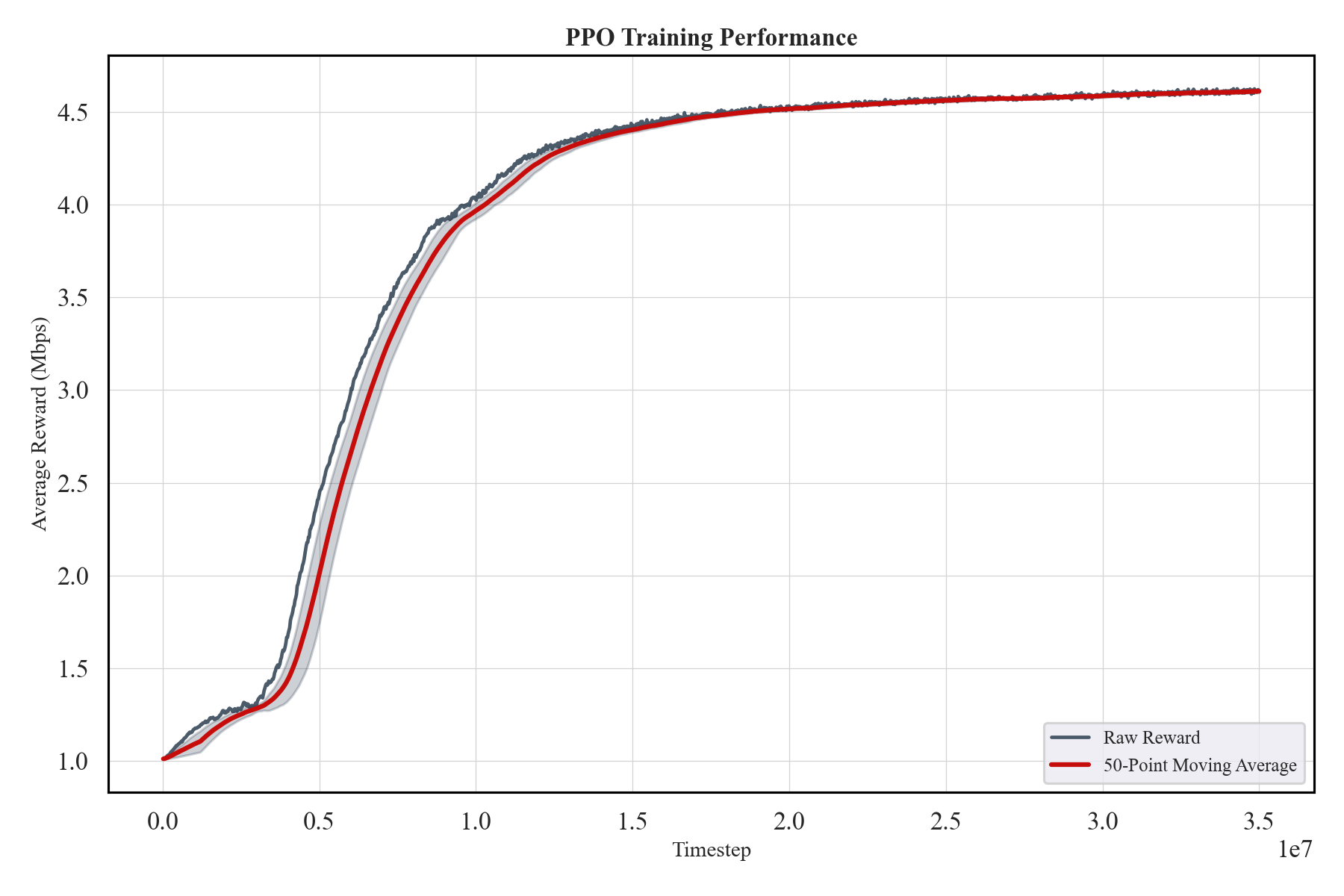}
\caption{MH-PPO training convergece.}
\label{fig:ppo_training_curve_convergence}
\end{figure}
In the simulation, we consider a segment of the SS DT of the Howard University campus in downtown Washington, DC, USA with an area of $330 \times 300$ m. The training and evaluation data are generated by our custom CT, which couples the \textit{Sionna} RT engine with the 3GPP antenna gain model. In contrast to simplified trajectory models that confine UAVs to predetermined circular or linear flight paths, our dataset generation procedure randomizes UAV positions within the three-dimensional volume of the considered drone corridor for each scenario. This approach produces richer channel diversity by yielding a diverse distribution of link distances, elevation angles, and LOS conditions, thereby preventing the agent from overfitting to repetitive geometric patterns while efficiently reflecting the stochasticity of real-world deployments. For each system configuration characterized by the number of UAVs $M$ and flight altitude $\xi$, we generate 8000 independent multi-UAV scenarios for training and 2000 scenarios for out-of-distribution testing. Each dataset incorporates the ray-traced CIR as a four-dimensional tensor of shape $\Xi \times M \times L \times N$, where $\Xi$ denotes the number of scenarios.

This proposed data-driven MH-PPO model is first trained offline and subsequently deployed in the live network for inference and testing. During training, channel gains are generated using a high-fidelity (HF) CT setup in \textit{Sionna} that models the CIR with $10^6$ rays. For online inference, the trained MH-PPO policy along with baselines are evaluated using channel gains obtained from an ultra–HF CT configuration with $10^8$ rays, corresponding to a two-order-of-magnitude increase in ray-tracing resolution and and DT accuracy in \textit{Sionna}. We assume this ultra-HF CT is the closest to real world channel measurements \cite{lee2025analysis}. While testing, entirely unseen scenarios were used for inference. This approach is critical for two reasons: first, it creates rich channel 
diversity by sampling a wide range of link distances and angles within the deployment area, preventing the agent from overfitting to specific geometric patterns. Second, it enables evaluation of generalization to 
unseen channel realizations and UAV configurations within the same SS environment, which is representative of operational scenarios where the network topology remains fixed but user distributions vary dynamically. It is worth mentioning that, all baselines and proposed methodologies were evaluated under identical conditions, i.e., environment and per-BS capacity constraints. All experiments are conducted on a Ubuntu 24.04 LTS workstation equipped with AMD Ryzen™ 9 9950X (16 cores, 32 threads, 5.7 GHz) processor, 128 GB RAM, NVIDIA RTX 4000 Ada Generation 20 GB Memory. We implemented the MH-PPO algorithm on Python 3.12 with PyTorch 2.8.
\begin{table}[t]
\centering
\caption{Summary of Baseline Methods}
\label{tab:baseline_summary}
\resizebox{\columnwidth}{!}{%
    \begin{tabular}{@{}lccc@{}}
    \toprule
    \textbf{Method} & \textbf{Category} & \textbf{Inference Complexity} & \textbf{CSI Required} \\
    \midrule
    MH-PPO (Proposed) & Learning & $\mathcal{O}(M \cdot LN)$ & Full \\
    DQN & Learning & $\mathcal{O}(M \cdot LN)$ & Full \\
    Two-Stage & Optimization & $\mathcal{O}(\max\{M, LN\}^3)$ & Full \\
    Max-Gain & Heuristic & $\mathcal{O}(M \cdot LN)$ & Partial \\
    Closest-BS & Heuristic & $\mathcal{O}(M(L + LN))$ & Partial \\
    Random & Naive & $\mathcal{O}(M)$ & None \\
    \bottomrule
    \end{tabular}%
}
\end{table}

\vspace{-1em}
\subsection{Results and Analysis}
We evaluate the proposed MH-PPO framework across nine distinct system configurations, spanning three UAV densities ($M \in \{20, 25, 30\}$) and three flight altitudes: $\xi \in \{60, 80, 100\}$~m. Intermediate altitude points (70~m and 90~m) shown in Figs.~\ref{fig:throughput_20uavs}-\ref{fig:throughput_30uavs} 
are obtained via Piecewise Cubic Hermite Interpolating Polynomial (PCHIP) interpolation for visualization continuity. All methods are tested on 2000 previously unseen scenarios generated with the ultra-HF CT ($10^8$ rays), ensuring that reported metrics reflect generalization to new channel realizations within the Howard University deployment area rather than in-sample memorization. This evaluation protocol validates the agent's ability to adapt to varying UAV distributions and channel conditions while maintaining the same SS propagation environment. This proposed inference scheme yields the predicted UAV-BS-beam association performance in terms Mbps. 

\subsubsection{Throughput Performance}

Fig.~\ref{fig:throughput_20uavs} presents per-UAV throughput as a function of altitude for $M = 20$. At $\xi = 60$~m, MH-PPO achieves 12.59~Mbps, representing a 720.6\% improvement over Closest-BS (1.53~Mbps) and a 43.8\% gain over DQN (8.76~Mbps). The Two-Stage Hungarian algorithm achieves only 3.61~Mbps despite its near-optimal signal-power objective, incurring a 248.7\% deficit relative to MH-PPO due to its neglect of interference during assignment. As altitude increases, all methods experience throughput degradation consistent with increased path loss. At $\xi = 100$~m, MH-PPO delivers 7.75~Mbps (408.1\% gain over Closest-BS), while its decay rate of 38.5\% from 60~m to 100~m remains comparable to DQN's 34.0\%, indicating consistent generalization across altitude variations.

Increasing UAV density to $M = 25$ intensifies resource competition and inter-UAV interference. As shown in Fig.~\ref{fig:throughput_25uavs}, MH-PPO achieves 11.25~Mbps at $\xi = 60$~m, corresponding to an 806.9\% improvement over Random and 279.5\% over Hungarian. The MH-PPO-DQN gap widens substantially in this configuration: MH-PPO outperforms DQN by 120.8\% (11.25 vs.\ 5.09~Mbps), suggesting that off-policy methods suffer from distribution mismatch as action space dimensionality scales with $M$. At $\xi = 100$~m, MH-PPO maintains superiority with 5.52~Mbps versus DQN's 3.89~Mbps, a 41.9\% margin demonstrating robust generalization to challenging propagation conditions.

The $M = 30$ configuration represents the most demanding regime, where the UAV-to-resource ratio approaches system capacity. Fig.~\ref{fig:throughput_30uavs} demonstrates that MH-PPO achieves 8.45~Mbps at $\xi = 60$~m, a 766.4\% improvement over Closest-BS and 116.2\% over DQN. This widening advantage validates the hypothesis that on-policy methods exhibit superior stability in high-dimensional, interference-limited environments. At $\xi = 100$~m, MH-PPO delivers 4.32~Mbps, maintaining gains of 321.5\% and 109.6\% over Closest-BS and Hungarian, respectively.

\begin{figure}[t]
\centering
\includegraphics[width=0.95\columnwidth]{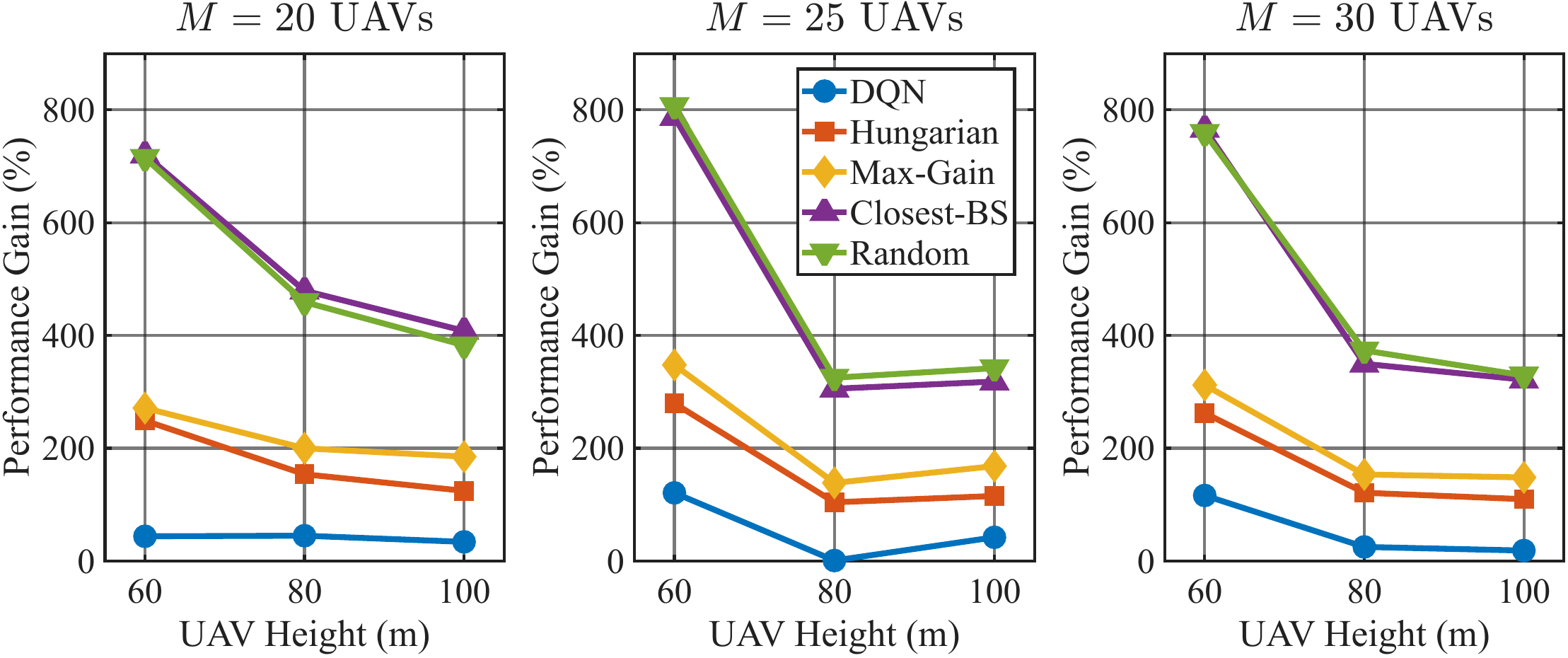}
\caption{Achieved performance gain over baselines}
\label{fig:ppo_gains}
\end{figure}

A consistent trend emerges across configurations: MH-PPO's relative advantage increases with UAV density. At $\xi = 60$~m, MH-PPO's gain over Hungarian grows from 248.7\% ($M = 20$) to 279.5\% ($M = 25$), reflecting the increasing value of interference-aware policies as the network approaches saturation. The Max-Gain heuristic underperforms Hungarian by 6–15\% due to its sequential processing bias, while Closest-BS and Random serve as performance floors, confirming that neither geometric proximity nor uniform selection provides viable allocation strategies. Fig.~\ref{fig:ppo_gains} consolidates performance gains across all configurations.


\subsubsection{Throughput Distribution Analysis}
Fig.~\ref{fig:cdf_subplots} presents CDFs of per-UAV throughput, providing insight into fairness and worst-case performance beyond aggregate metrics. At $M = 20$ and $\xi = 60$~m, MH-PPO's 5th percentile throughput reaches 6.05~Mbps compared to 1.87~Mbps for DQN, a 3.2$\times$ improvement in tail performance. Note that, the $5^{th}$ percentile represents the worst-case user performance or "cell-edge" throughput. In this circumstance, 95$\%$ of UAVs achieves at least 6.05~Mbps which ensures a minimum quality of service. Furthermore, this gap indicates that MH-PPO not only maximizes aggregate throughput but also provides substantially better guarantees to UAVs experiencing unfavorable channel conditions. The DQN distribution exhibits a long left tail with approximately 10\% of UAVs receiving throughput below 2~Mbps, reflecting the instability of off-policy learning. Hungarian produces a compact distribution (5th to 95th percentile: 2.89–4.39~Mbps) due to its deterministic nature, but this consistency comes at the cost of a limited throughput ceiling. Table~\ref{tab:worst_case_throughput} summarizes the 5th percentile throughput across all configurations, quantifying the floor performance that each method guarantees to disadvantaged UAVs.

 \begin{table}[htbp]
    \centering
    \caption{5th Percentile Throughput (Mbps): Worst-Case UAV Performance}
    \label{tab:worst_case_throughput}
    \begin{tabular}{cc|cccc}
        \toprule
        $M$ & $\xi$ (m) & MH-PPO & DQN & Hungarian & Max-Gain \\
        \midrule
        \multirow{3}{*}{20} & 60  & \textbf{6.05} & 1.87 & 2.89 & 2.68 \\
                            & 80  & \textbf{5.55} & 3.03 & 2.82 & 2.26 \\
                            & 100 & \textbf{4.66} & 2.13 & 2.59 & 2.13 \\
        \midrule
        \multirow{3}{*}{25} & 60  & \textbf{4.30} & 0.93 & 2.37 & 2.08 \\
                            & 80  & \textbf{3.63} & 0.92 & 2.18 & 1.76 \\
                            & 100 & \textbf{3.27} & 0.91 & 2.14 & 1.66 \\
        \midrule
        \multirow{3}{*}{30} & 60  & \textbf{3.78} & 0.66 & 1.98 & 1.71 \\
                            & 80  & \textbf{3.31} & 0.86 & 1.82 & 1.54 \\
                            & 100 & \textbf{3.11} & 0.74 & 1.71 & 1.43 \\
        \bottomrule
    \end{tabular}
\end{table}

Moreover, as density increases, distributional differences become more pronounced. At $M = 25$ and $\xi = 60$~m, MH-PPO maintains a 5th percentile of 4.30~Mbps while DQN degrades to 0.93~Mbps. This 4.6$\times$ gap reveals that DQN increasingly fails to coordinate associations under heightened interference. The high-density configuration ($M = 30$, $\xi = 60$~m) amplifies this trend: MH-PPO's 5th percentile (3.78~Mbps) exceeds DQN's by 5.7$\times$, and remarkably surpasses DQN's median throughput (3.91~Mbps). DQN's extreme variance at $M = 30$ and $\xi = 100$~m (5th to 95th percentile: 0.74–18.44~Mbps) indicates policy degeneration into high-variance behavior where some UAVs receive excellent service while others are effectively denied connectivity.

\subsubsection{Training Convergence}

Fig.~\ref{fig:ppo_training_curve_convergence} illustrates MH-PPO's learning dynamics over $3.5 \times 10^7$ timesteps for $M = 30$ and $\xi = 80$~m configuration. The training curve exhibits three characteristic phases: initial exploration (0–0.5$\times 10^7$ steps) where rewards approximate the Random baseline, rapid improvement (0.5–2.0$\times 10^7$ steps) with monotonic reward increase from 1.5 to 3.5~Mbps, and fine-tuning (2.0–3.5$\times 10^7$ steps) where the policy converges to approximately 4.5~Mbps. The smooth ascent without oscillations or collapses validates MH-PPO's stability in high-dimensional action spaces. The entropy coefficient annealing (0.2 to 0.005) appropriately balances exploration and exploitation, while GAE-based advantage estimation enables continued improvement in later training stages.

\begin{figure}[t]
\centering
\includegraphics[width=01\columnwidth]{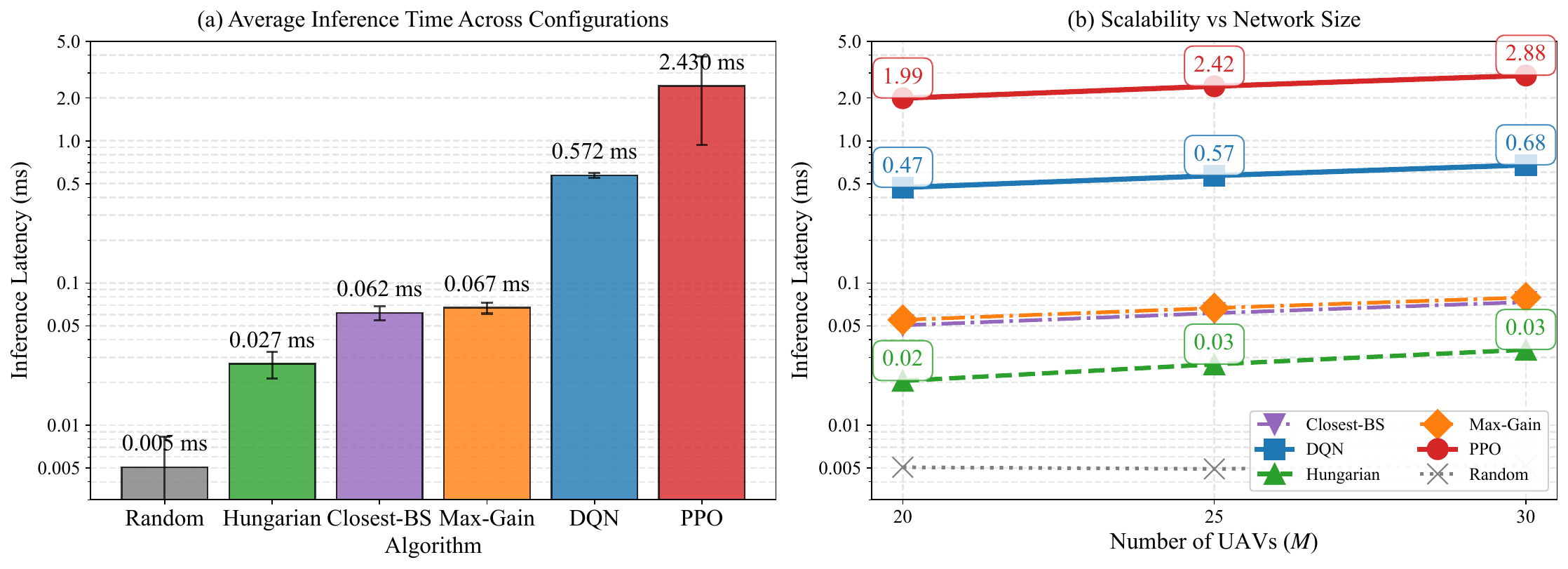}
\caption{Inference over live network performance}
\label{fig:inference_analysis}
\end{figure}

\subsubsection{Inference Analysis on Live Networks}
Real-time operation is critical for UAV resource allocation, where decisions must be computed within the channel coherence time to ensure CSI validity. For UAVs traveling at 20-50~km/h, the coherence time $T_c \approx \frac{9}{16\pi f_m}$ ranges from 10-100~ms \cite{Goldsmith2005}, where $f_m = vf_c/c$ is the maximum Doppler frequency. Additionally, 3GPP-compliant aerial systems require Command-and-Control packet latencies below 40~ms for flight stability \cite{bordin20255g}.

Fig.~\ref{fig:inference_analysis}(a) shows mean inference latencies averaged across all nine configurations ($M \in \{20, 25, 30\}$ UAVs, $\xi \in \{60, 80, 100\}$~m). MH-PPO achieves $\bar{\tau}_{\text{inf}} = 2.43$~ms ($\sigma = 1.51$~ms), compared to DQN's 0.57~ms and Hungarian's 0.027~ms. This latency represents only 2.4\% of the minimum coherence time ($T_c^{\min} \approx 10$~ms) and is $16.5\times$ faster than Command and Control (C2) packet requirements, ensuring decisions execute before channel conditions evolve. MH-PPO trades 1.86~ms additional latency versus DQN for 42.6\% throughput improvement (12.48~Mbps vs. 8.75~Mbps), demonstrating favorable performance-latency balance. Moreover, MH-PPO outperforms Hungarian by 109\% in throughput while incurring only 2.40~ms overhead, proving that learned policies achieve both superior spectral efficiency and real-time operation.

Fig.~\ref{fig:inference_analysis}(b) validates computational scalability. As $M$ increases from 20 to 30 (50\% growth), MH-PPO latency rises from 1.99~ms to 2.88~ms (44.7\% increase), confirming sub-linear $\mathcal{O}(M \cdot LN)$ complexity. Each of $M$ actor heads processes shared features through independent $(128 \to 64)$ linear layers, yielding total FLOPs proportional to $M \cdot (128 \cdot 64 + 64 \cdot LN)$. In contrast, exhaustive search over $(LN)^M = 64^{30} \approx 10^{54}$ joint actions is computationally infeasible. DQN exhibits similar scaling (0.47~ms to 0.68~ms) while Hungarian's cubic complexity ($\mathcal{O}(M^3)$) becomes apparent at 0.020~ms to 0.033~ms. The sub-linear scaling ensures continued feasibility for ultra-dense deployments, with extrapolated $\tau_{\text{inf}} \approx 4.2$~ms for $M=100$ UAVs still within coherence time constraints.

These results establish that our proposed DT-driven MH-PPO enables millisecond-granularity resource allocation for next-generation SS aerial networks, compressing intractable combinatorial optimization into real-time neural inference while maintaining superior throughput performance.
\section{Conclusion}
\label{sec:conclusion}
In this paper, we developed a novel DRL framework to enable real-time, coordinated UAV–BS–beam association while mitigating interference in a DT-enabled aerial corridor. Our approach leverages a ultra-HF CT, constructed from SS ray-tracing data, to generate a large-scale offline dataset. A PPO agent, featuring a scalable multi-head actor-critic architecture, was trained on this dataset to learn a direct mapping from complex channel states to resource allocation decisions, with network constraints enforced through the reward function. We conducted a comprehensive performance evaluation, benchmarking the trained PPO agent against classical heuristics, a near-optimal optimization method, and a baseline DQN agent on unseen test scenarios. The simulation results concluded that the proposed MH-PPO framework consistently achieves the highest average network throughput, significantly outperforming all baselines. Altogether the agent demonstrated strong generalization to unseen channel realizations within the deployment area and achieved state-of-the-art performance with millisecond-level inference latency, a key requirement for real-time applications. This work validates the efficacy of DTs for (a) training DRL algorithms to solve complex RRM problems and (b) enabling a coordinated RRM scheme while addressing challenges associated with global CSI acquisition. Future work could explore the integration of UAV mobility into the state space, investigate multi-agent reinforcement learning techniques for decentralized control,  cross-domain transfer learning, and comprehensive comparison with advanced off-policy methods.

\balance
\bibliographystyle{IEEEtran}
\bibliography{references_TeamB}
\balance

\end{document}